\documentclass[%
 reprint,
nofootinbib,
nobibnotes,
 amsmath,amssymb,
 aps,
prl,
floatfix,
]{revtex4-2}
\usepackage{lipsum}

\usepackage{afterpage}
\usepackage{hyperref}
\usepackage{graphicx}
\usepackage{dcolumn}
\usepackage{bm}

\usepackage{algorithmic}
\usepackage[linesnumbered,lined,algoruled]{algorithm2e}

\usepackage[dvipsnames]{xcolor}
\newcommand\green[1]{#1}

\makeatletter
\def\maketitle{
\@author@finish
\title@column\titleblock@produce
\suppressfloats[t]}
\makeatother

\begin{document}

\author{Matan Yah Ben Zion$^{1,2,3}$}
\thanks{To whom correspondence should be addressed.\\
E-mail:  matanbz@gmail.com}

\author{Jeremy Fersula$^{1,2}$}
\author{Nicolas Bredeche$^{2}$}

\author{Olivier Dauchot$^{1}$}
\affiliation{$^{1}$Gulliver UMR CNRS 7083\\
\normalsize{ESPCI, PSL Research University, 75005 Paris, France}}
\affiliation{$^{2}$Institut des Syst\`{e}mes Intelligents et de Robotique\\
\normalsize{Sorbonne Universit\'{e}, CNRS, ISIR, F-75005 Paris, France}}
\affiliation{$^{3}$School of Physics and Astronomy, and the Center for Physics and Chemistry of Living Systems\\
\normalsize{Tel Aviv University, Tel Aviv 6997801, Israel}}

\title{Morphological computation and decentralized learning in a swarm of sterically interacting robots 
}
\homepage[This manuscript has been accepted for publication in Science Robotics. This version has not undergone final editing. Please refer to the complete version of record at \href{https://www.science.org/doi/10.1126/scirobotics.abo6140}{https://www.science.org/doi/10.1126/scirobotics.abo6140}. The manuscript may not be reproduced or used in any manner that does not fall within the fair use provisions of the Copyright Act without the prior, written permission of AAAS.]{}

\begin{abstract}
Whereas naturally occurring swarms thrive when crowded, physical interactions in robotic swarms are either avoided or carefully controlled, thus limiting their operational density. Here we present a mechanical design rule that allows robots to act in a collision-dominated environment. We introduce the Morphobots --- a robotic swarm platform developed to implement embodied computation through a morpho-functional design. By engineering a 3D-printed exoskeleton we encode a re-orientation response to an external body force (such as gravity) or a surface force (such as a collision). We show that the force-orientation response is generic, and can augment existing swarm-robotic platforms (e.g Kilobots) as well as custom robots even 10 times larger. At the individual level, the exoskeleton improves the motility and stability, and also allows to encode two contrasting dynamical behaviors in response to an external force or a collision (including collision with a wall or a movable obstacle, and on a dynamically tilting plane). This force-orientation response adds a mechanical layer to the robot's sense-act cycle at the swarm level, leveraging steric interactions for collective phototaxis when crowded. Enabling collisions also promotes information flow, facilitating online distributed learning. Each robot runs an embedded algorithm that ultimately optimizes collective performance. We identify an effective parameter that controls the force-orientation response and explore its implications in swarms that transition from dilute to crowded. Experimenting with both physical swarms (of up to 64 robots), and simulated swarms (of up to 8192 agents) show that the effect of morphological-computation increases with growing swarm size.
\end{abstract}

\maketitle

\section*{Introduction}


A robotic swarm is a synergetic ensemble capable of a greater task than its constituents~\cite{beni1993swarm}. In a swarm, self-organization is an emergent property, making the swarm resilient to the malfunction of an individual. Swarm engineering poses an interdisciplinary challenge, combining electrical and mechanical engineering, computer science, and non-equilibrium statistical physics ~\cite{Brambilla2013,Hamann2018}. Discovering robust design rules for a swarm offers an opportunity to simplify the specifications of the individual robotic units, reducing manufacturing complexity and making production at-scale accessible ~\cite{Brooks1989,dorigo2020reflections,floreano2021individual,dorigo2021swarmPastPresentFuture}. Recent development in robotic swarms demonstrated potential applications in collective construction~\cite{Werfel2014}, coordinated motion for flying UAVs~\cite{Hauert2008,viragh2014flocking,vasarhelyi2018optimized,mcguire2019minimal}, patrol in open water~\cite{gomes2016cooperative} or underwater exploration~\cite{zahadat2016division,berlinger2021implicit}. These applications are characterized by swarms where robots explicitly avoid physical contact~\cite{schranz2020swarmapplis}. Another approach for swarm design is to keep a cohesive swarm, where robots are constantly touching, keeping a physical interaction that allows continuous transmission of information~\cite{Saldana2017,Li2019, Savoie2019, Du2020, Li2021, Ozkan-Aydin2021}. This approach showed success in self-assembly and morphogenesis~\cite{Rubinstein2014,Slavkov2018,Wang2021} as well as coordination of a multi-cellular robotic body~\cite{li2019particle,oliveri2021continuous, Deblais2018, Boudet2021}. To date, artificial swarms are designed to operate exclusively in a dilute, collision-avoided setting, or a cohesive dense population. Resorting to in silico experiments to guide swarm design in a collision-dominated swarm proves limited by the ability to effectively account for the full mechanical interaction. There is a need for physical design rules and robust algorithms where robots maintain a high degree of autonomy in solitude, but also retain functionality in a crowded environment. 

Unlike artificial robotic swarms, naturally occurring swarms show excellent flexibility in both dilute and highly dense environments throughout the animal kingdom. At the cellular scale, an individual bacterium can explore space by performing a simple run-and-tumble~\cite{Berg2000}, but can also form a dense, highly active bacterial colony that displays internal turbulent flow~\cite{Wensink2012} thereby mixing in nutrients~\cite{Xu2019, Costerton1999}. On the millimeter scale, an individual larva can search for food through chemotaxis but self-organize into a living fountain when a large localized food source is present~\cite{Shishkov2019}. Macroscopic animals such as fish and birds can individually nest and mate but can come together into giant schools and flocks to evade predation~\cite{Procaccini2011,BBC2015}. Even pedestrians flow can emerge from simple body collisions~\cite{Moussaid2012CrowdDisaster}. Swarm cooperation is found on a range of length scales, suggesting that only simple primitive behaviors are required and physical interactions can be used for the swarm's advantage~\cite{Bechinger2016}. Interestingly, self-organization in dense assemblies is not a prerogative of living beings. It has been shown that collective motion can emerge from independent mindless robots physically interacting with one another, considering a stripped-down flavor of swarm robotics as a particular kind of active matter with no explicit decision computed at the level of the robots ~\cite{Hara2003,deseigne2010collective,Vicsek2012,lam2015self,Bricard2013,boudet2021collections}. 

Our motivation is to expand the roboticists' toolbox by taking advantage of a generic mechanical response: the tendency to re-orient in response to an external force. We then demonstrate the role of this morphological response in different robotic tasks, both for an individual robot, and a robotic swarm. We introduce two designs to show that this force-orientation response can come in different flavors, aligning with, or against an external force. We show that this force-orientation response can also be used in a robotic swarm platform, giving control over the outcome of a robot-robot collision and the resulting collective behavior.

To achieve this we augment robots with a flexible exoskeleton and show that the morphological difference is encoded in a single effective parameter, \green{labeled $\kappa$}, that controls the outcome of collisions. We show two different designs that have similar dynamics for free individual robots (i.e a single robot moving on a smooth, flat surface, without experiencing any external force, nor touching any obstacles or robots). But when experiencing an external force, or collision, the two designs have an opposite response. We show that the passive dynamics of the design allow tuning the robot's response to align or oppose an external force. Upon collision, a robot can push against, or slide along an obstacle or another robot, based solely on its mechanical design. The emerging collective dynamics add a morphology layer to a robot's computation, facilitating a combined physical and logical swarm architecture~\cite{pfeifer2006body,pfeifer2009morphological}. We quantitatively analyze the response to an external force and find empirically that it can be described by a single effective parameter, $\kappa$. We show experimentally, theoretically, and numerically that a change in the sign of $\kappa$ inverts the response to an external force, making a robot go down or up a force field. We find $\kappa$ to be a key design parameter in the response of the robots when encountering an obstacle (be it a wall or another robot). Using positive and negative $\kappa$ makes two primitives that qualitatively alter the swarm's dynamics in a collision-dominated, crowded environment. We also show that $\kappa$ can be used to guide the design of collective transport. 

\begin{figure}[htbp]
\centering
\includegraphics[width=1\columnwidth]{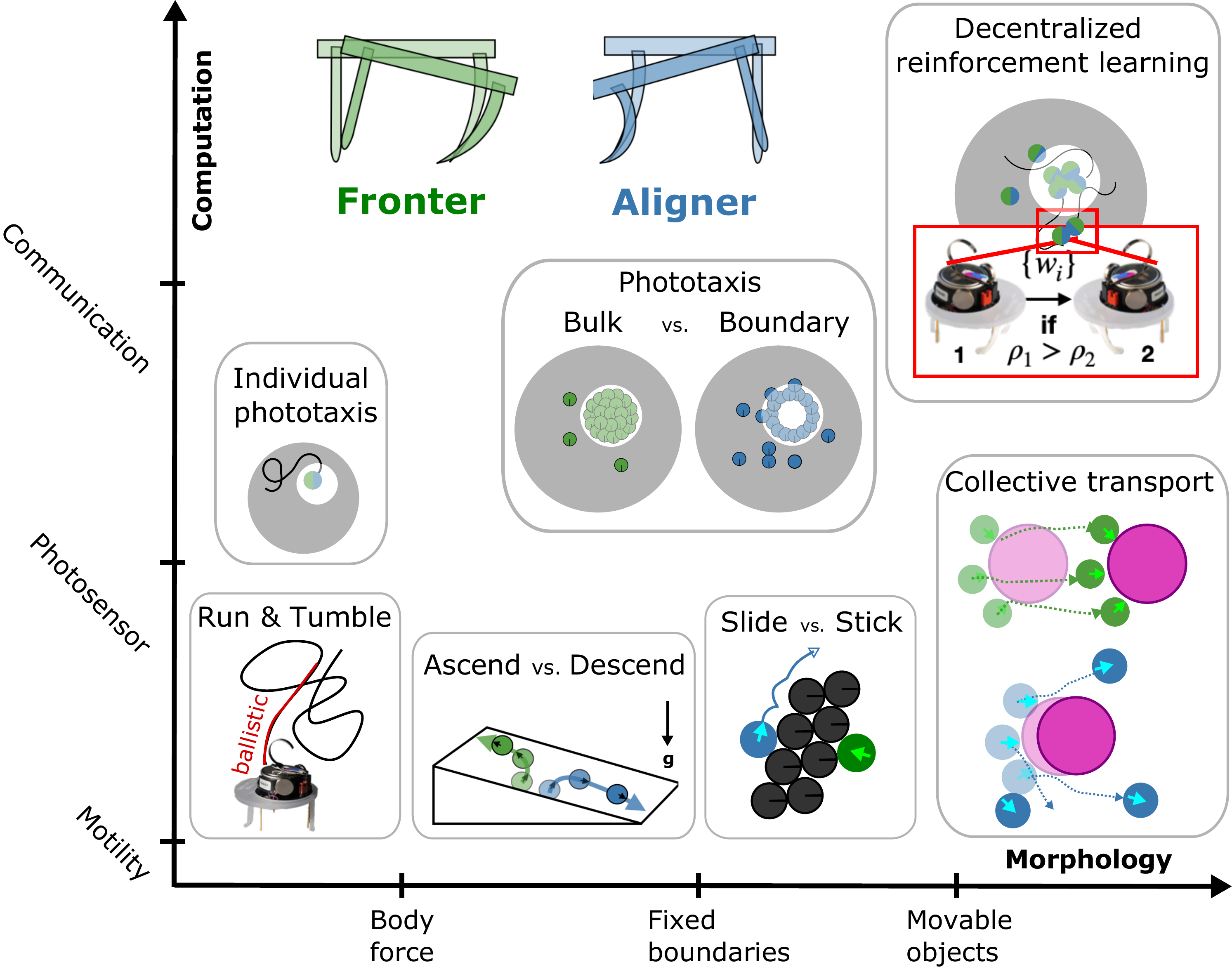}
\caption{{\bf Article at a glance}. Swarm robotics design space spans both morphology and computation. Experiments with 3D-printed flexible exoskeletons show they can direct the outcome of collisions between robots. The mechanical design of the exoskeleton, fronter versus aligner, controls whether a robot ascends or descends a hill, which translates to pushing against or sliding along a wall. The exoskeletons' morphological computation extends the ability of the swarm to execute collective tasks, including transport, phototaxis, and decentralized learning.}
\label{fig:MorphoComp}
\end{figure}

We then add a computational layer, where in parallel to the implicit morphological computation, robots execute an explicit sense-act cycle. We experimentally study phototaxis in a partially lit arena, setting a swarm of $N=64$ robots to search for the lit region. We set up the illumination field to be unoriented (coming directly from above), and without local light gradients, thus ensuring robots cannot use local information to guide their way to the lit region. Moreover, the lit region is designed to be too small to host the whole swarm. This restriction forces the robots to physically interact, turning the phototaxis task into an exercise of collective aggregation with sterical interactions, going from dilute (in the dark) to crowded (in the light). Studying swarms with positive and negative $\kappa$ reveals a qualitative and a quantitative different collective phototaxis, directly attributed to the nature of this effective parameter.

Finally, we show that the swarm can collectively learn to execute the phototactic task when operating in an environment where individual robots transition between dilute and crowded regions. We implement a decentralized reinforcement learning algorithm inspired by social learning~\cite{Watson2002a,bredeche2018ee,fontbonne2020, bredeche2022rspt}, and find that when collisions are allowed the swarm fluidizes, and effectively converges to a successful phototactic strategy despite the sparse and intermittent communication network. We further support our experimental observations with agent-based model simulations, finding that morphology is central to collective behavior and becomes increasingly important with growing swarm size. The interplay of morphology and computation is graphically illustrated in Fig.~\ref{fig:MorphoComp}, summarizing the findings of our work.


\section*{Individual Dynamics}
\subsection*{The Morphobots}
\begin{figure*}[htbp]
\centering
\includegraphics[width=0.95\textwidth]{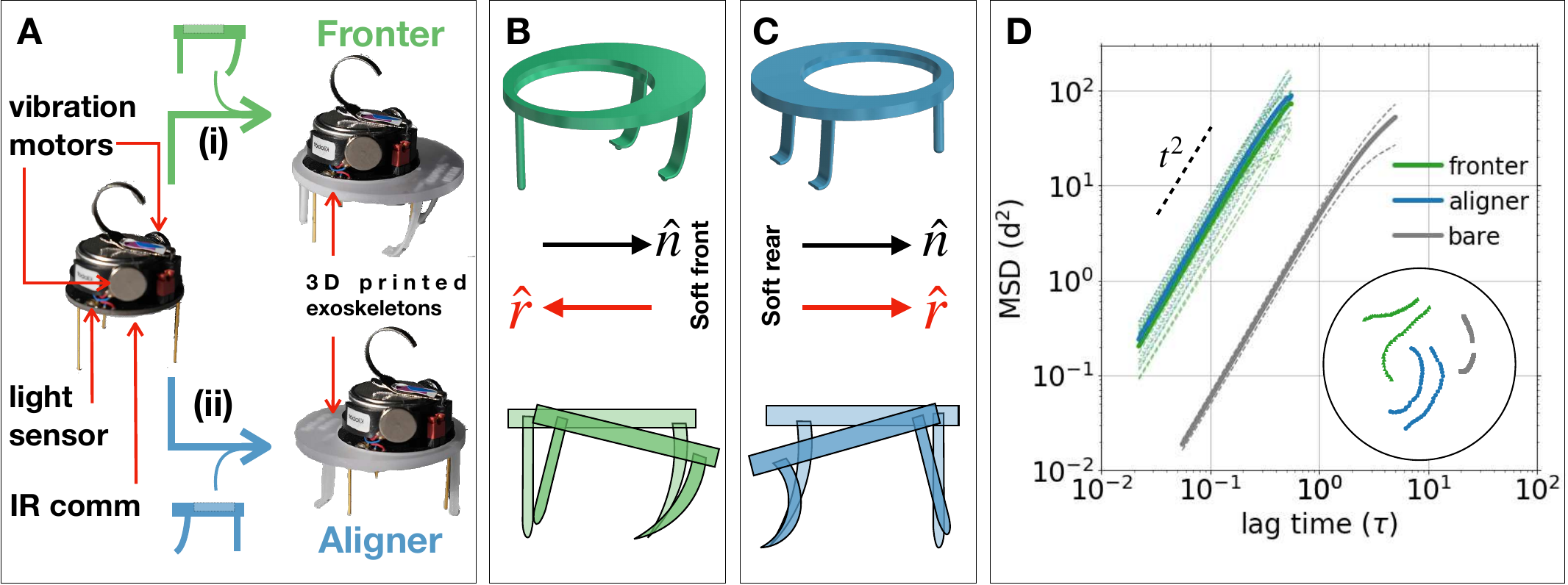}
\caption{Augmenting Kilobots with exoskeletons makes Morphobots.  {\bf A} a Kilobot with a light sensor, IR communication, and two differential drive vibration motors, is fitted into one of two 3D printed exoskeleton designs - (i) fronter, and (ii) aligner. Both fronter {\bf B} and aligner {\bf C} exoskeletons are tripods, but the restitution vector, $\hat{r}$ of the aligners (fronters) is parallel (anti-parallel) to the orientation vector $\hat{n}$. {\bf D} mean square displacement versus time shows the ballistic response for both designs ($\propto t^2$), with a typical speed of 5.2 cm/s (4.8 cm/s) for the aligners (fronters). Also shown, the mean square displacement of a bare Kilobot which after calibration can also move ballistically, but at a speed one order of magnitude slower (0.53 cm/s). Inset: 12-second long trajectories of fronters (green) and aligners (cyan), and a 36-second long trajectory of bare Kilobots (gray) inside the 150 cm diameter arena.}
\label{fig:Morphobot}
\end{figure*}

We make Morphobots by augmenting Kilobots~\cite{Rubinstein2014} with a 3D-printed exoskeleton. The Kilobot is a $3.4$~cm tall, $3.3$~cm diameter unit, standing on 3 rigid legs made of thin metallic rods. The Kilobots are capable of differential drive locomotion using two vibrators. When either is activated, a Kilobot turns at about $45\deg$ per second, and when both are activated, the Kilobot moves forward. The power source is a rechargeable lithium-ion battery that can keep the Kilobot running for a few hours. Kilobots are equipped with an infrared transmitter and receiver so that they can communicate with each other. The transmitter of a robot sends light toward the surface which reflects up to the receiver of another nearby robot. Obtaining straight motion over long distances is difficult and requires lengthy calibration of the individual vibrators \cite{Rubenstein2013}, moreover, the calibration itself drifts over time.

To both improve the robot's nominal speed, and eliminate the need for individual calibration, we encapsulate the Kilobots within exoskeletons that change how collision and locomotion are performed.  We made two tripod-based designs (a pair of flexible legs and an opposing stiff leg), with both designs having a circular chassis (diameter $d=4.8$cm). The two designs, which we shall call fronter versus aligner, differ in the fore-versus-aft positioning of their flexible legs (see Fig.~\ref{fig:Morphobot}-A,B,C and Methods). The natural vibration of the semi-flexible legs is tuned near resonance with the vibration motors, to maximize the coupling with their drive. Using an anisotropic flexible leg design (1mm thick, 6 mm wide, and 20 mm long, see Supplementary Materials), vibrations are coupled predominately to the forward stick-slip motion of the robots. The exoskeleton does not interfere with the ability of the robots to communicate, and has multiple roles including energy storage of vibrations, near-resonance coupling to motors, anisotropic design to translate motor-vibrations into a forward motion. The exoskeleton allows a Morphobot to operate at an order of magnitude higher speed than the Kilobots (see Fig. \ref{fig:Morphobot}-D and Supporting Movie 1). More importantly, it allows control of the response to external forces and collisions.

By tracking the motion of individual Morphobots, we find that aligners and fronters show near-identical motility. Motility is defined by the magnitude of the displacement of an individual robot over different timescales, and measured by tracking the time evolution of their two-dimensional coordinates, $\vec{r}\left(t\right) = \left(x\left(t\right), y\left(t\right)\right)$, when running in the experimental arena~\cite{Allan2019}. Only a low concentration of robots was used (less than 1\% filling fraction) excluding the effect of collisions, and robots were tracked only when away from the walls of the arena. The mean square displacement, $\rm{MSD} = \langle \Delta r^2\left(\Delta t\right) \rangle_N$, of an ensemble of noisy walkers, presents two dynamical regimes~\cite{Fodor2018}. On short time intervals $\Delta t$, the displacement of the robots is aligned with the direction of their individual orientation and displacements are proportional to $\Delta t$. The motion is said to be ballistic when robots go roughly on a straight line at a constant speed: $\langle \Delta r^2\left(\Delta t\right) \rangle_N = v_0^2 \Delta t^2 $, where $v_0$ is the instantaneous speed.

On long time intervals, the noise has reoriented the robots several times and displacement grows as $\sqrt{\Delta t}$ as robots explore the space randomly. The motion is now said to be diffusive $\langle \Delta r^2\left(\Delta t\right) \rangle_N = 4 D_0 \Delta t $, where $D_0$ is called the diffusion constant. The noisy trajectories of the robots originate from the stochastic nature of their vibrational drive \cite{Scholz2016}, as well as the roughness of the arena and internal electronic drift. The two regimes ($\propto \Delta t$ and $\propto \sqrt{\Delta t}$) are separated by a crossover, which takes place on a time scale called the persistence time, $\tau_p$ (see below), to which corresponds a persistence length $\langle \Delta r^2\left(\tau_p\right) \rangle_N^{1/2}$. Fig.~\ref{fig:Morphobot}-D displays the mean square displacement measured for a set of Morphobots equipped with the two types of the exoskeleton, together with their ensemble average (the MSD of 8 robots of each design were measured, shown as separate curves in Fig.2D, see also Table \ref{tbl:alignersFronters} ). One sees that the difference in motility between the two designs is no larger than the robot-to-robot variation within each design. The observed ballistic nature of the mean square displacement  ($\propto v_0^2 \Delta t^2$) indicates that the persistence length of the motion is larger than the size of the circular arena (diameter $D=150$ cm). Fitting the ensemble mean square displacement in the ballistic regime, we find that the mean nominal speeds are $5.2\pm 0.9$ cm/s and $4.8\pm 0.1$ cm/s for the aligners and fronters respectively. The measured effective parameters are summarized in Table \ref{tbl:alignersFronters}.

Altogether, the anisotropic leg design allows us to use higher motor drive while keeping the robot's trajectories straight, alleviating the need for individual robot calibration, and resulting in a nominal speed of $v_0\simeq5$ cm/s, a ten-fold increase, relative to the bare Kilobot design (see Supporting Movie 1). The notion of using passive-dynamic machines~\cite{Collins2005} allowed to achieve a similar feat in bristle bots, using rows of soft legs~\cite{Giomi2013}, and in the commercially available {\it HEXBUG} \cite{Deblais2018, Dauchot2019, Boudet2021, Baconnier2022}. Using the differential drive, robots can also pivot at $45\deg $ per second. 
Programming alternated straight and pivoting motion of random duration and random direction, we can set the Morphobots into a run-and-tumble motion~\cite{Tailleur2008}, where, on average, the robots run straight during a run time $\tau_{run}=2$ s, with a velocity to be specified, and tumble during a mean tumble duration $\tau_{tumble}=4$ s. The total duration of a run-and-tumble sequence sets $\tau_{RT} = \tau_{run}+\tau_{tumble}$, with the resulting persistent time being $\tau_P \approx 18\;\rm{s}$ (See SM).

\subsection*{Fronters vs. Aligners response to an external force} 

\begin{figure*}[t]
\centering
\includegraphics[width=0.85\textwidth]{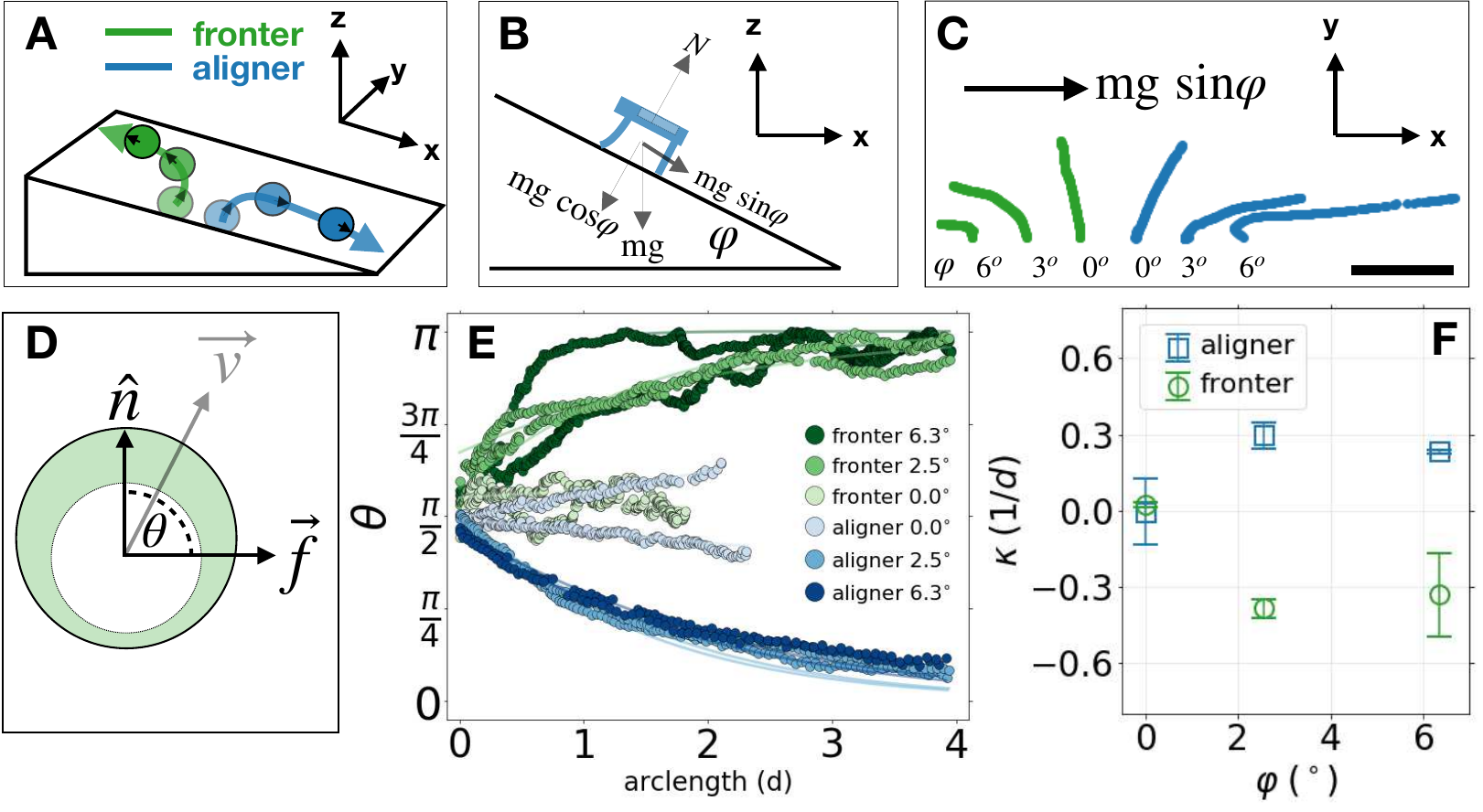} 
\caption{The orientation response of aligners and fronters to an external force is equal and opposite.  {\bf A} Sketch of the experimental setup for measuring the dynamical response on an inclined plane (see Movies 1, 3-6). {\bf B} Force diagram of a robot placed on an inclined plane. For small slopes ($\varphi \le 6^\circ$), robots experience an external body force. {\bf C} Trajectories of aligners (cyan) and fronters (green) moving on different slopes ($\varphi$).  Aligners and fronters have an opposite response to the external force - aligners' orientation turns downwards as they quickly descend, while fronters turn upwards, and slowly ascend.  {\bf D} Throughout its motion, the orientation of the robot, $\hat{n}$, forms an angle $\theta$ with the external force, $\vec{f}$, together setting the net velocity $\vec{v}$ (see Eq. \ref{eqVelocityDif}). {\bf E} Tracking the orientations shows that aligners turn downhill ($\theta \rightarrow 0$), aligning with the external force, while fronters turn uphill ($\theta \rightarrow \pi$), fronting the external force (see also Movies 3-6). Solid curves are fits to Eq. \ref{eqOrientation}, from which $\kappa$ can be extracted (see SM). {\bf F} \green{Average fitted $\kappa$ for each incline angle and robot-design (error bars are the standard error).} For finite tilt angle, the response of the two designs is approximately equal and opposite $\kappa_{aligner} = -\kappa_{fronter} \approx 0.3/d $. At a slope near zero, the response is dominated by the robot's internal bias (left/right curving and internal noise \cite{Scholz2016}), but averages to zero. Scale bar is 20 cm.}
\label{fig:slope}
\end{figure*}

We systematically measure the response of the two Morphobots designs to an external body force by placing the robots on a plane tilted by a small angle, $\varphi$. The two designs have a polar different response: aligners go downhill, while fronters climb against gravity (Fig.~\ref{fig:slope}A,B and Supporting Movies 1-6). The effect is consistent, regardless of the initial orientation of the robot being along the positive $y$ direction (downhill to its right), or negative $y$ direction (downhill to its left), excluding the effect of an internal left/right bias (see Movies 3-6). We systematically increase the slope (see Fig.~\ref{fig:slope}C), and find that with increasing $\varphi$, the trajectories turn increasingly curved, until the robot orients downhill (aligner) or uphill (fronter). 

To quantify the different responses of the robots to an external body force, we modify a model introduced in the context of active matter to describe self-aligning particles~\cite{Dauchot2019}. A Morphobot has two translational $\vec{r} = (x,y)$ and one orientational $\hat{n}= (cos(\theta),sin(\theta))$ degree of freedom. 
In the overdamped limit, the deterministic dynamics of the velocity $\vec{v}$ and orientation $\hat{n}$ of a Morphobot with a nominal speed $v_0$ subjected to a body force $\vec{f}$ (see Fig. \ref{fig:slope}D) obey the following equations :
\begin{eqnarray}
\frac{d\vec{r}}{dt} &=& \vec{v} = v_0 \hat{n} + \mu \vec{f}
\label{eqVelocityDif}\\
\frac{d\hat{n}}{dt} &=& \kappa \left(\hat{n}\times\vec{v}\right) \times\hat{n},
\label{eqOrientationDif}
\end{eqnarray}
where $\mu$ is the particle mobility and $\kappa$ is the \green{alignment parameter} --- an effective parameter describing the re-orientation by force. The first equation simply states that the velocity is the additive combination of the propulsion velocity $v_0 \hat{n}$ and the speed induced by the external force, $\mu \vec{f}$. The second equation contains the key ingredient of the model. It describes the reorientation of the Morphobot along the direction of its motion. The coupling between an external force and the orientation is captured through an effective \green{alignment parameter}, $\kappa$. This parameter measures to what extent an external force, $\vec{f}$, affects the orientation of a robot,  $\hat{n}$. In the most simple case, $\kappa = 0$, and an external force does not affect the orientation of a robot (Eq. \ref{eqOrientationDif} becomes redundant). However, generically force and orientation do couple, and $\kappa$ is finite ($\kappa \neq 0$). Moreover, $\kappa$ is signed as it can be positive or negative. This means that when subjected to an external force the robot does not only translate, but it also rotates. The larger the magnitude of the \green{alignment parameter}, $\|\kappa\|$, the quicker a robot's orientation will respond to an external force. There are two important types: when positive ($\kappa>0$) {\it a robot aligns parallel with an external force}. Here we call robots with positive $\kappa$ aligners. When negative ($\kappa<0$) {\it a robot aligns anti-parallel to an external force}. Here we call robots with negative $\kappa$ fronters. $\kappa$ is an effective parameter and results from the design details of the robot (material properties, mass distribution, drive mechanism, and so on). And yet $\kappa$ is empirically well defined as it can be experimentally measured and quantitatively compared between different robotic platforms. For a constant force in the $\hat{x}$ direction $\vec{f} = f \hat{x}$, with $f = mg\rm{sin}\phi$, where $m$ is the mass of the Morphobot and $g$ is the acceleration of gravity, the dynamical evolution of the orientation, $\theta=\theta\left(t\right)$ relative to the $x$ axis obeys a closed form equation :
\begin{equation}
\frac{d\theta}{dt} = -\kappa \mu f \rm{sin}\theta.
\label{eqPendulum}
\end{equation}
Eq. \ref{eqPendulum} is the equation for an overdamped simple pendulum, which can be solved directly to give $\rm{tan}\frac{\theta}{2} = e^{- \kappa \mu {\it f} t}$ (for initial condition $\theta\left(t=0\right)=90^o$). In the limit where $v_0\ll\mu f $, when the orientation is expressed as a function of the arclength, $\theta \left(s\right)$, it is found to be independent of the external force and mobility (see Supplementary Materials),
\begin{equation}
\theta \left(s\right) = 2 \rm{atan}\left(e^{-\kappa s}\right).
\label{eqOrientation}
\end{equation}
%

To find the alignment parameter, $\kappa$, we tracked the orientation of individual robots as they move across a fixed inclined plane (see Fig. \ref{fig:slope}E, Movies 3, 4, and Methods). In the absence of an external force (zero slope, $\phi=0^{\circ}$), both fronters and aligners move on roughly straight lines. For steeper inclines ($\phi =3^{\circ}$ and $\phi=6^{\circ}$), both designs show a similar rate for the convergence of the orientation. Fitting the evolution of the orientation to Eq. \ref{eqOrientation}, we extract the alignment parameter, $\kappa$. The magnitude of $\kappa$ for both designs is similar, but with opposite signs: $-\kappa_{fronter} \approx \kappa_{aligner} \approx 0.06\;\text{cm}^{-1}$. The measured effective parameters are summarized in Table \ref{tbl:alignersFronters}.

\begin{figure}[t]
\centering
\includegraphics[width=\columnwidth]{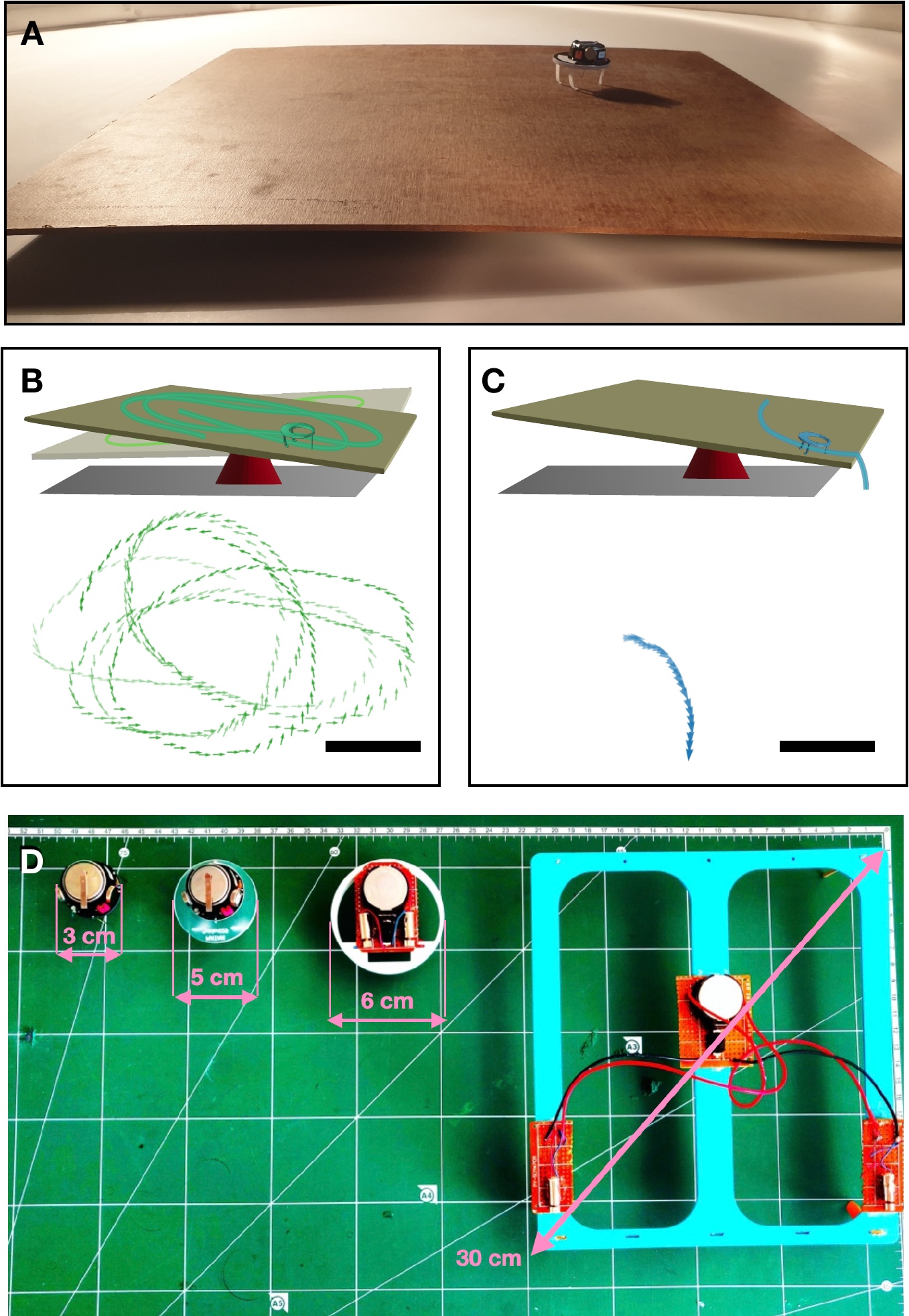} 
\caption{Fronters and aligners respond differently to a dynamic environment. {\bf A} A Morphobot on a swinging table setup for testing morphological response to a dynamically changing environment. {\bf B} A fronter continually turns uphill, moving in a circular motion while staying on the swinging table (see also supporting Movie 2), whereas an aligner ({\bf C}) quickly turns downhill and falls off. Top shows schematics, bottom shows measured trajectories. ({\bf D}) The design rules apply also to custom-made robots (not Kilobot based), producing the expected force alignment response with Morphobots as much as 10 times larger than a Kilobot (see supporting Movies 5,6). Scale bars are $10\;\rm{cm}$.}
\label{fig:seesaw}
\end{figure}

We find that the design rules controlling the alignment parameter apply also in a dynamically changing environment as well as to robotic platforms of different sizes (see Fig.~\ref{fig:seesaw}). Placing a Morphobot of either type on a seesaw gives rise to a unique behavior (Fig. \ref{fig:seesaw}): An aligner turns quickly downhill and falls off the seesaw, while a fronter turns upwards and climbs. Once passing the center of the seesaw, the mass of the robot suffices to tip the plane of the swinging table. The fronter is now facing downhill, yet, again, as it moves, it turns around and recommences climbing, until passing the center of the seesaw once more, where these dynamics repeat, effectively tracing loops (see Fig. \ref{fig:seesaw}B and supporting Movie 2). Moreover, the design rules for the Morphobot generalizes to a custom robotic platform independent of the Kilobots (Fig. \ref{fig:seesaw}D) and can be made 10 times larger (see Movies 5 and 6).

\begin{table}[b]
\centering
\begin{tabular}{c c c} 
                & $v_0 \left(d/s\right)$   & $\kappa \left(1/d\right)$ \\ 
                \hline
    Aligners    & $1.1 \pm 17\%$           & $0.3 \pm 10\%$\\  
    Fronters    & $1.0\pm 2\% $            & $-0.3 \pm 20\%$\\
 
\end{tabular}
\caption{Measured nominal speed and alignment parameter of aligner and fronter designs.}
\label{tbl:alignersFronters}
\end{table}f

\subsection*{Interaction with a wall} 

\begin{figure}[t]
\centering
\includegraphics[width=1\columnwidth]{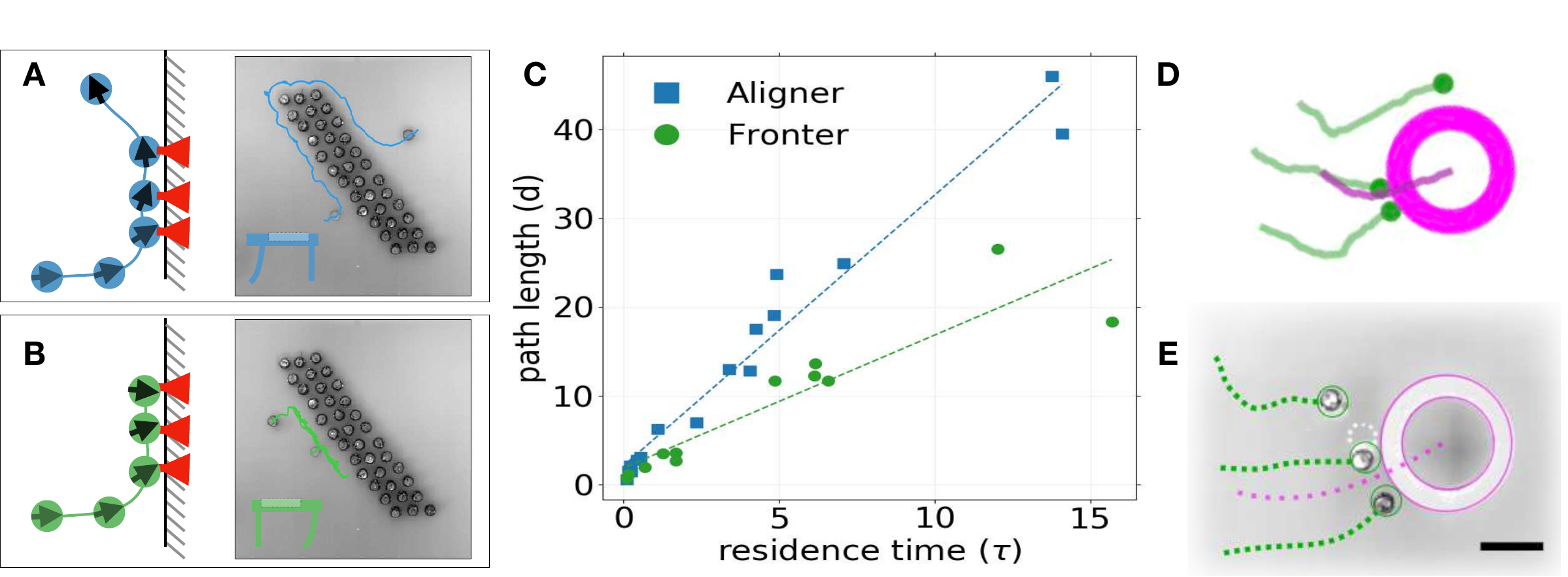} 
\caption{Tuning interaction with a static and a moving obstacle using morphological computation. The morphological interaction of the Morphobots with walls can be explained using their response to an external force. When a Morphobot collides with a wall, each impingement can be seen as a momentary normal force.  {\bf A} Aligners turn to align their orientation (black arrow) with the normal force (red). Once aligned along the wall, there is no further rotation as there are no more collisions, and the robot slides along. {\bf B} When a fronter collides with a wall, its orientation (black) turns anti-parallel with the normal force (red), and the robot continues to push into the wall. When programmed to run-and-tumble, aligners and fronters interact differently with a wall of stationary robots where aligners ({\bf A}) cover a much larger distance along the wall relative to fronters ({\bf B}). {\bf C} Multiple collision experiments show that aligners and fronters spend similar time near a wall ($3 \tau_{p}\approx 1\;\rm{min}$) yet aligners move twice as fast. {\bf D} Brownian dynamics simulations predict that active particles with a negative alignment parameter, $\kappa <0$ (fronters), push a movable object. {\bf E} With real robots: placing a round movable obstacle shows that a swarm of fronters is capable of transporting an object. Scale bar is $10\;\rm{cm}$.}
\label{fig:wall}
\end{figure}

Using the response of the Morphobots to an external force we can predict their steric interaction with stationary walls, movable objects, and with one another. Morphological interaction with a boundary (stationary or movable) can be seen as a series of collisions, at each of which the robot experiences an external force (see Fig. \ref{fig:wall}-A). With each collision, a Morphobot slightly re-orients. An aligner orientation approaches the normal force from the wall (but leaves when parallel to the wall). A fronter turns against the normal force (and towards the wall). This is confirmed by the direct observation of Morphobots sent to move towards a wall of stationary robots (see Fig. \ref{fig:wall}-B). \green{Arranging turned off robots into three, tight rows forms effective wall with which a moving robot mechanically interacts.} On average, both designs spend similar time near the wall ($\sim3\tau_{p}\approx 1\;\rm{min}$) but aligners travel a path along the wall which is twice longer than that of the fronters (see Fig. \ref{fig:wall}-C and Movie 1). Using a multi-agent Brownian dynamics simulation of active soft discs, we were able to predict the outcome of collisions based on the alignment strength parameter $\kappa$. During a collision, an agent with $\kappa>0$ (aligner), turns towards the normal force, and away from the obstacle, whereas an agent with $\kappa<0$ (fronter) turns against the normal force and into the obstacle thus pushing it. 
Also, simulating agents with $\kappa <0$ near a passive particle shows that the agents collectively move the passive object (see Fig. \ref{fig:wall}-D). This has been confirmed experimentally by using a group of fronters that work together to push a circular object (see Fig. \ref{fig:wall}-E and Supporting Movie 1).
Previous work showed that collective transport can be implemented using global control \cite{Shahrokhi2016}. Here we extend this notion and show that a single, effective \green{alignment parameter}, $\kappa$, captures embodied dynamics of sterically interacting agents, reducing the required computing resources for simulating a large swarm in a crowded environment.

\green{
\section*{Collective Dynamics}

We turn to study the ability of a swarm of Morphobots to execute a collective task, focusing on phototaxis, where robots explicitly use their sensory and computational power in addition to the morphological interaction. Phototaxis is a common behavioral trait found in nature where an organism changes its motion in response to locally measured light. Phototaxis pervades the animal kingdom, from individual bacterium to groups of vertebrates. Phototaxis is classified as either positive, where the organism seeks to be present in the light (such as moths \cite{Lees2019a}) or negative, where the organism prefers the dark (such as golden shiners \cite{Berdahl2013}). Here we test the ability of a robotic swarm to perform this widespread natural behavior, with an emphasis on the collective ability to maximize the time spent in the light (positive phototaxis). We program the robots to change their speed in response to the locally measured light intensity (also known as photokinesis~\cite{NULTSCH1971}) and show that the robots' morphology can be used to improve their collective phototaxis when crowded. Stimulus-induced locomotory movements (or {\it taxis}) constitute an important building block in swarm robotics~\cite{sahin2004swarm,Brambilla2013,Bayindir2016,Hamann2018,Savoie2018,Savoie2019}, and several bio-inspired algorithmic implementations have been proposed for phototaxis using ambient light sensors~\cite{hamann2008spatial,schmickl2009getintouch,kernbach2009honeybee,arvin2011imitation,li2019particle}.}

\subsection*{Collective phototaxis in the presence of steric interactions}
\begin{figure*}[htbp]
\centering
\includegraphics[width=1\textwidth]{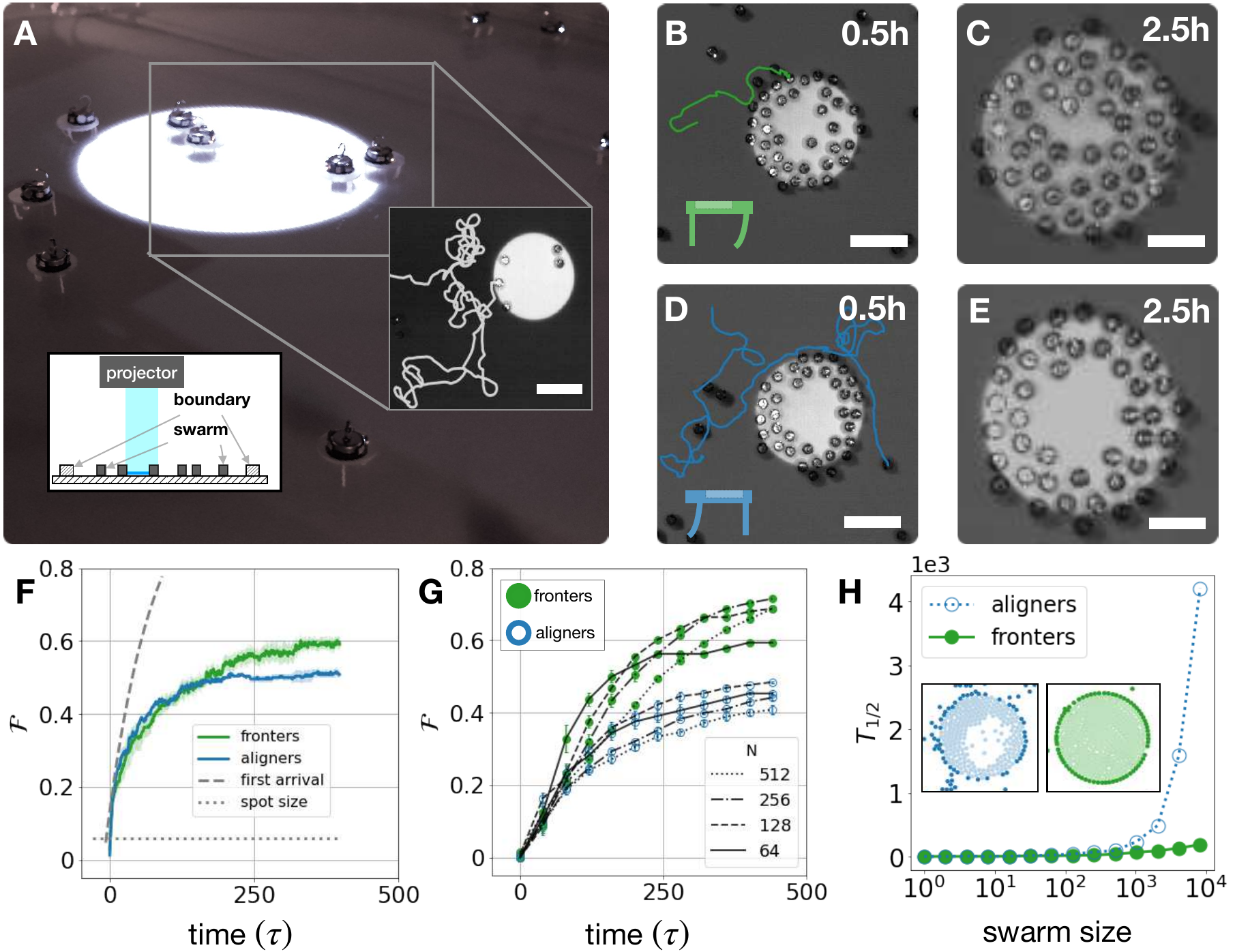} 
\caption{
The collective ability of a robotic swarm to phototaxis is constrained by steric interactions. {\bf A} A swarm of phototactic robots is placed in an arena patterned with an un-oriented light field. As there are no local gradients to guide a robot to the light source, robots perform a simple run-and-tumble in the dark and stand in the light. {\bf B} When a fronter arrives at the light spot, it collides with robots already taxied at the perimeter, but progressively pushes its way in. Repeating this process results in the complete coverage of the light spot ({\bf C}). When an aligner arrives at the light spot ({\bf D}) it slides along the wall of robots and leaves. This leads to only partial filling of the light spot ({\bf E}), leaving an empty void in the center.  {\bf F} The resulting fraction of swarm in the light as a function of time shows that at early times ($\tau \approx 30 \tau_p \approx 10\;\rm{mins}$), both design show equal performance, quantitatively consistent with a diffusion-limited reaction (dashed line). At later times, fronters outperform aligners (average of 4 realizations with error bars being standard deviation). {\bf G} Simulating swarms with over two orders of magnitude more agents show a sharp slowdown in the phototaxis of aligners with swarm size. {\bf H} The typical time for half of the swarm to enter the light $T_{1/2}$ grows rapidly for an increasingly larger swarm of aligners, while having only a small effect on a swarm of fronters of similar size. Scale bars are 20 cm.}
\label{fig:morphoswarm}
\end{figure*}

\green{The experimental setup for collective phototaxis was designed to prevent 
 an individual robot from accessing global information that could guide its sense-act cycle.} For this, we set a swarm ($N=64\;\rm{bots}$) of either aligner ($\kappa>0$) or fronter ($\kappa<0$) to explore a circular arena (diameter $D=150$ cm) with a lit spot that covers only $\sigma=6\% $ of the area of the arena (diameter $b=36\;\rm{cm}$, see Fig. \ref{fig:morphoswarm}-A and Section \ref{fig:arena} in SM). The light is unoriented (coming from above), and there are no local light intensity gradients to guide a robot to the lit region (the light field is essentially flat). This excludes the use of traditional phototactic algorithms such as steepest descent search~\cite{schmickl2011beeclust,Wang2021} or reorientation~\cite{Rubinstein2014}. The lit region can not accommodate more than 80\% of the robots (about 50 Morphobots), challenging the swarm to transition between sparse and crowded.

As a first step, we consider ad hoc calibrated swarms, where the Morphobots' policies are defined as follows. Each Morphobot continuously monitors local light intensity: when in the dark (below a predefined intensity threshold, $P_{th}$) the robot performs a run-and-tumble motion, with a running velocity $V_0$. When entering a bright region (light intensity above $P_{th}$), the robot stops, $V_1=0$. 
At early times, when there are only a few robots in the lit region, the two designs show similar performance as measured by the fraction of the swarm that is in the light 
\begin{equation}
\mathcal{F} \equiv \frac{N_1}{N},
\label{eqSwarmInLight}
\end{equation}
where $N_1$ is the number of bots taxied in the light. At the beginning of every experiment, robots are placed in the arena with random positions and orientations. At early times, $\mathcal{F}$ is quantitatively captured as a diffusion-limited process (see Fig. \ref{fig:morphoswarm}-F), with the kinetic reaction constant found using Smoluchowski first passage computation for diffusive particles with an effective diffusion constant, $D_{eff} \approx 4.2^2 \;cm/s$ (See Supplementary Materials)~\cite{Smoluchowski1916}.  After approximately 10 minutes, both swarms deposit the first layer of robots at the perimeter of the lit region (Fig.~\ref{fig:morphoswarm}B, D). From this point onward, the phototaxis rate slows down and deviates from the simple diffusion-limited reaction, as crowding of taxiing robots at the perimeter of the light spot physically prevents further robots from entering the lit region (Fig.~\ref{fig:morphoswarm}C, E).  This steric effect is quantitatively more pronounced for the aligner design, as can be seen for the overall performance given by the fraction of the swarm that has entered the lit region over long times (Fig.~\ref{fig:morphoswarm}F).

The reason is that the collective success of each swarm depends on the morphological interactions of a robot impinging from the dark region on the first layer of robots at the perimeter of the lit region. Individual Morphobots arriving at the wall of stationary robots show qualitatively different dynamics for either design: an aligner ($\kappa >0$) that arrives at the wall tends to reorient along the wall until it finally leaves, whilst a fronter Morphobot ($\kappa <0$) tends to ``dig in'', pushing already taxied robots into the lit region until taxied itself (see Fig. \ref{fig:morphoswarm}B, D). These individual events add up: fronting Morphobots progressively push the perimeter inwards, constantly filling the bulk of the lit region with robots. By contrast, the aligning Morphobots keep reorienting away from the robots at the perimeter, leaving a hollow in the lit region (see Fig.\ref{fig:morphoswarm} C, E and Supporting Movie 1). This discrepancy exacerbates as the swarm size increases. By simulating up to 8192 phototactic agents that follow Eqs. \ref{eqVelocityDif},\ref{eqOrientationDif} we find that a swarm of aligners ($\kappa >0$), becomes considerably slower to phototaxis relative to an equivalent swarm of fronters ($\kappa <0 $). The typical time, $T_{1/2}$, to have half of the swarm in the light ($\mathcal{F} = 1/2$) grows much faster for aligners than for fronters (see Fig.~\ref{fig:morphoswarm}G, H).  The above findings demonstrate that the collective ability of the swarm to perform a phototactic task at scale hinges on a subtle morpho-functional difference.


\subsection*{Distributed learning of a phototactic strategy}

The swarms considered thus far had the environmental information hard-coded into individual robots before deployment --- the sense-act cycle of every robot had a pre-programmed, ad-hoc knowledge of the correct light threshold for phototaxis (which is an unrealistic assumptions when robots are to be deployed in an unknown environment). Concurrent challenges however require deployment of robotic systems in unknown terrains~\cite{sahin2004swarm,Brambilla2013,Bayindir2016,Hamann2018}, including extra-terrestrial exploration, underwater surveying, and search-and-rescue missions~\cite{Uckert2020,Fiorello2020,Berlinger2021}. Under such circumstances, robots will lack an a priori model of the environment that can be pre-programmed. This requires an added layer to the robots' controller for online tuning of the sense-act cycle based on continually acquired knowledge. In such conditions, a distributed robotic system offers robustness with scale, improving both exploration and survival. A distributed system however also poses a challenge as the swarm has to learn collectively.

\begin{figure*}[t]
\centering
\vspace{-0.5cm}
\includegraphics[width=0.95\textwidth]{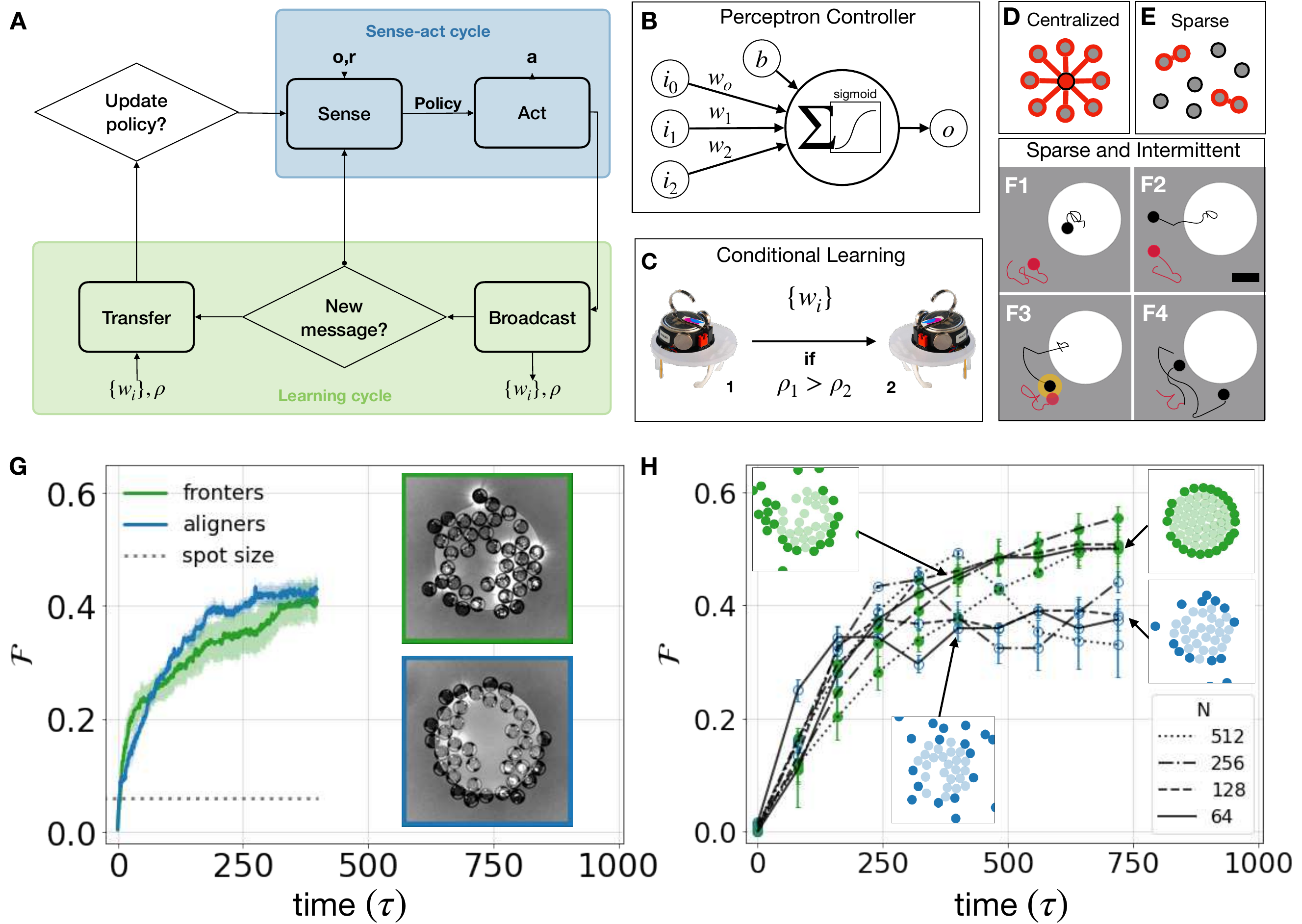} 
\vspace{-0.5cm}
\caption{
Distributed online reinforcement learning algorithm and deployment of a swarm of Morphobots. {\bf A} The embedded algorithm for updating the phototactic policy of each robot, is composed of two main moduli: a sense-act cycle and a learning cycle. {\bf B} The sense-act cycle is encoded into a Perceptron: external stimuli are weighted and summed, evaluated through a sigmoid function, the output of which is binarized to select an action. The weights, $w$, define the policy $\pi$ of the robot. {\bf C} The learning cycle is based on the conditional diffusion of the policies across the population, biased by a reward function $\rho$ evaluated by each robot: when robot 2 receives a message containing the policy control parameters $\{w_i\}_1$ and the reward $\rho_1$ of robot 1, robot 2 will inherit the new policy, if it is associated with a higher reward $\rho_1>\rho_2$. {\bf D} Illustration of a centralized communication network where all agents (gray circles) are connected to a central operator (red circle). {\bf E} An illustration of a sparsely connected network where agents communicate only when in close proximity. {\bf F1}-{\bf 4} Illustration of a sparse and intermittent communication where a robot (black curve) already in the light source ({\bf F1}) departs ({\bf F2}) \green{and broadcasts within communication range (yellow circle)} its reward and policy  to a robot in the dark (red curve) ({\bf F3}), which inherits the new policy ({\bf F4}). Note that the broadcasting robot may have a wrong policy, effectively spreading false information, with neither robot knowing what is the best policy for phototaxis. {\bf G} A swarm of Morphobots with randomly initialized light thresholds, collectively learns its environment, increasing the swarm fraction in the light. Results are the average of 4 runs with error bars being the standard deviation. Insets show that fronters (top) and aligners (bottom) develop a different spatial arrangement in the highlighted region. {\bf H} Evolution of the swarm fraction in the light for multi-agent simulations of swarms of different sizes learning to phototaxis using the distributed on-line, reinforcement learning. At earlier times, swarms of fronters and aligners develop the spatial arrangement found in the experiments. At later times (inaccessible experimentally) a swarm of fronters compactly arrange inside the light spot a larger fraction of the swarm in the light compared to the swarm of aligners (see also insets).}

\label{fig:Learning}
\end{figure*}

In this final section we test whether a swarm can collectively learn even when robots collide. We show that in a swarm of Morphobots, both execution and learning can be made fully distributed (see Fig.~\ref{fig:Learning}). We then proceed to test the effect of the two designs on the collective performance and find numerically that over long enough time, a swarm of fronters out-performs a swarm of aligners. Since the robots are allowed to collide, the swarm is fluidized~\cite{Sole2019}. An individual collision event may be stochastic, but the mean data flow within the swarm becomes predictable. Unlike a solid or a jammed state (where individuals do not move), here the swarm acts as a fluid, meaning that it satisfies good mixing, as robots frequently exchange neighbors and visit various parts of the arena, sustaining a flow of information (see Figs. \ref{fig:IndividualDynamics}, \ref{fig:LargeLightSpot} in supporting information). Each robot evaluates its performance and compares it to other encountered robots. The selection algorithm can be made distributed by implementing a decentralized online reinforcement learning scheme, to make the swarm adaptive. 

Previous work showed that a swarm can learn a phototactic strategy with a policy defined by the weights of an Artificial Neural Network as a controller, $\pi = \{ w_i \} $~\cite{trianni2003evolving,dorigo2004aggregation,soysal2007aggregation}. There, the learning was implemented using an evolutionary algorithm to act as a direct policy search, using a reinforcement learning method, incrementally refining behavioral policies by directly tuning the policies' parameters~\cite{Nolfi2000,Doncieux2015}. The swarm is randomly initialized with a pool of sense-act policies, and in an iterative selection process, only the best-performing policies are carried over to the next generation. In these algorithms, the execution is indeed distributed over the swarm, but it is a central computer performing selection and optimization (see Fig.~\ref{fig:Learning}D). Such schemes are impossible to implement outside the lab if the swarm has no access to an external operator for guidance, nor a global communication channel to compare performances. In what follows we show that despite having a sparse and intermittent network, communication in a swarm of Morphobots 
becomes effectively connected across time thanks to the swarm’s fluidity, facilitating decentralized learning (see Fig.~\ref{fig:Learning}F).

%

For a successful reinforcement learning algorithm to converge, a suitable reward function, $\rho$, must be defined~\cite{kaelbling1996reinforcement}. As an input, the reward function receives the robot's sensory information $\rho = \rho\left(\{i_i\}\right)$ and outputs a scalar score (see Fig. \ref{fig:Learning}A and Algorithm \ref{algoVanillaHIT} in SM). The combination of an individual robot sense-act policy, $\pi = \pi\left(\{w_i\}\right)$, and the environment it sampled, results in a robot's reward at a given time. Robots can also communicate when within communication range and exchange their policies and rewards (Fig. \ref{fig:Learning} C, F1-F4). Following such an exchange, a robot can choose to adopt the received policy based on its associated reward, operating under the assumption that a higher reward is the result of a superior policy. Unlike classical evolutionary algorithms that use a centralized computer to compare performance, the scheme presented here allows embedding both the evaluation and the selection within the robot. In other words, robots collectively perform a kind of social learning, where successful individual innovations diffuse over the population~\cite{Watson2002a,bredeche2018ee, bredeche2022rspt}. Note that the assumption that a higher reward is a result of a superior policy may not be always true. From time to time robots with an incidental favorable history can have a disproportional higher reward. Such bots will propagate a poor policy globally to the swarm, effectively spreading `fake news'. This local-to-global gap challenges the chosen reward function to be robust to such fluctuations. The fluid nature of the swarm of Morphobots allows us to choose $\rho$ that overcomes this culprit. Most swarm robotic systems employ a collision avoidance algorithm that slows down their collective dynamics, jamming the information flow, and creating deadlocks~\cite{Werfel2014}. By contrast, the Morphobots are allowed to collide, sustaining good mixing, and offering a quantitative link between global and local information (see Figs. \ref{fig:IndividualDynamics}, \ref{fig:LargeLightSpot} in Supplementary Materials). 

To formulate a proper reward function we note that the collective phototaxis, $\mathcal{F}$, as defined in Eq. \ref{eqSwarmInLight} is an ensemble average, $\langle \rangle_N$, counting the fraction of the swarm that is in the light at a given time. $\mathcal{F}$ can not be directly optimized by an individual robot as it stores global information. However in a system with good mixing, a time average can approximate an ensemble average $\phi\equiv \langle \rangle _T \approx \langle \rangle_N$. A time average equivalent to the ensemble average given in Eq.~\ref{eqSwarmInLight} is the mean time an individual robot spent in the light, $\Phi\equiv T_{{\rm in\,light}}/T \approx \mathcal{F}$. In a distributed robotic system with good mixing, a reward function that estimates the mean time the robot spent in the light can serve as a proxy for the global performance of a swarm of its kind,  $\rho \propto \Phi \approx \mathcal{F}$. Robot $i$ can evaluate the mean time it spent in the light through a temporal average of its light sensor signal:
\begin{equation}
\rho^i = \frac{1}{M}\sum^{M-1}_{k=0} P^i_k,
\label{eqReward}
\end{equation}
where $M$ is the number of averaged measurements, and $P^i_k$ is the the $k$'th measurement taken at constant time intervals. Since the time a robot spends in the light is inversely proportional to its speed in the light, $\Phi \propto \rho \propto 1/V_1$ the above choice of reward function is analytically linked to a given policy. This link allows us to show that when formulated as a dynamical process, collective learning will indeed converge on average (see SM for detailed derivation). 

To test the validity of the proposed decentralized learning scheme, we deployed a swarm of Morphobots in the same light-patterned arena, but this time with {\it randomly initialized policies} (See Fig.~\ref{fig:Learning}, Supporting Movie 1). A ``correct'' policy corresponds to a set of weights $\{w_i\}$ of the perception, where the robot slows down in the light. A ``wrong'' policy corresponds to a set of weights, where the robot does not slow down in the light. When randomly distributed in the large arena used (150 cm), robots are on average over 16 cm apart (see for example 01:41 in Movie 1). This is more than twice the maximum communication distance supported by the Kilobot platform (7 cm). At such a separation, many of the robots are out of communication range from one another making the network sparse. Not only the network is sparse, but the network is also intermittent. Even when two robots do come within communication range, they quickly move away. On average, the subgroup of the swarm that interacts at a given time makes up only a small fraction of the swarm, and this subgroup is constantly changing. Altogether making a sparse and intermittent communication network. We also program the robots to start with random initial policies by drawing their weights at equal probability from values ranging between 0-255 (see Methods Section and Supplementary Materials for details). At very early times only a few robots slow down in the light --- most robots either stand in the dark or move quickly through the lit region. And yet the swarm fraction in the light progressively increases with time. We tested swarms of both aligners and fronters, and found a consistent increase in the fraction of the swarm in the light reaching approximately $\mathcal{F} \approx 0.4$. At these swarm fractions in the light, the difference between fronters and aligners is not expected to be significant as is evident here (compare with Fig.~\ref{fig:morphoswarm}F at early times). The learning swarms had 8 times more robots in the light than the case of random placement that is, the fraction of the lit region, $\sigma = 0.06$. Given that Kilobots are ten times slower, and diffusion scales as the speed square ($\propto v_0^2$, see SI), a swarm of bare Kilobots would have taken 100 times longer to converge, making collective learning impossible. The fact that the adaptive swarm performed almost, but not quite as well as, the pre-calibrated swarm is not surprising and may be attributed to the fact that dynamics of about half of the randomly initialized population are very slow (as they stand in the dark), and given enough time, they will also eventually learn. This may be resolved by expanding the learning space of the sense-act cycle, to have the robots also learn the speed in the light, $V_1$.

We find that the difference between learning swarms of aligners and fronters increases with time. By extending the multi-agent numerical engine, we tested the learning process over a longer duration (inaccessible experimentally) and for larger swarms (up to 8 times larger, see Fig.~\ref{fig:Learning}H and Fig.~\ref{fig:simulationSnapshotsLearning} in supporting information). Implementing Algorithm \ref{algoVanillaHIT} numerically, we find that after approximately $\tau \approx 300-400$ from their initialization, swarms of either aligners or fronters reach the same fraction of robots in the light ($\mathcal{F} \approx 0.4$, Fig.\ref{fig:Learning}G), consistent with Morphobots studied in the laboratory. In both experiments and simulations, we find that at these intermediate times, the swarm of fronters tends to occupy the lit region while aligners remain on the perimeter (see Fig.~\ref{fig:Learning}G, H). Simulations show that at later times ($\tau>400$) this qualitative difference grows: the fraction of aligners in the light remains the same, whereas the swarm of fronters continues to crowd the lit region, reaching $\mathcal{F}\approx 0.55$. This effect is consistent with growing swarm size, showing the significance of morphological interaction at scale in adaptive swarms.

\section*{Conclusion}
In this research article, we introduced a robotic swarm that can operate in both dilute and dense settings, where collisions between robots occur frequently. We show that the robots' physical morphology plays an important role in terms of collective behaviors, where a seemingly small change in the shape leads to very different outcomes. Applying tools from statistical physics of active-matter to swarm robotics, we show that behavior found in natural swarms (such as ant colonies~\cite{Feinerman2018}) including collective self-aggregation and transport can be attained with limited use of software-controlled behaviors, by capitalizing on the morphological properties of robots, demonstrating morphological computation at work within the swarm.

We also showed that distributed online evolutionary reinforcement learning can be implemented in a dense robot swarm, to exploit or mitigate physical contingencies after deployment in the open. We find that this kind of artificial social learning, when robots locally exchange information about their behavioral strategies, can be modeled as a conditional diffusion process.

Future collective learning implementations in swarm robotics can go beyond the diffusion of behavioral strategies. An extreme case is of sparse swarms, where individual robots or groups may be temporarily disconnected from the whole~\cite{tarapore2020sparerobotswarms}. While dense and sparse robot swarms may be considered separately, a robot swarm in the open will likely have to switch from one configuration to another, requiring versatile social learning capabilities. A natural extension can be motivated by cultural evolution found in nature~\cite{whiten2021burgeoning}, which is built on the reformulation and combining previously learned behaviors to continuously discover behaviors of growing complexity.

Future physical models of a learning swarm should go beyond the mean-field approximation used here and account for the spatial density fluctuations (as is done for describing the motility-induced phase separation observed in active fluids~\cite{Tailleur2008,BenZion2022}). From that point of view, our work shows that building upon the recent theoretical progress made in describing the population of active particles, one can envision a rapid development of a statistical mechanics description of smart active matter, for future swarm engineering.

\vspace{-0.2cm}
\section*{Acknowledgements}
We thank Y. Lahini and N. Oppenheimer, and S. Elul. {\bf Funding}: This work was supported by the MSR project funded by the Agence Nationale pour la Recherche under Grant No ANR-18-CE33-0006. M.Y.B.Z. gratefully acknowledges support from the Ministry of Aliyah and Immigrant Absorption of Israel. {\bf Author contributions}: 
O.D, N.B, and M.Y.B.Z conceived the research. M.Y.B.Z Designed the experiments, theoretical models, and analysis. M.Y.B.Z and J.F performed the experiments and the data analysis. M.Y.B.Z, O.D, and N.B wrote the manuscript. {\bf Competing interests}: The authors declare no competing interests. {\bf Data and materials availability}: All (other) data needed to evaluate the conclusions in the article are presented in the article or the Supplementary Materials.

\section*{Data Availability}
All data and code required for reproducing the results in the manuscript are found in online repositories~\cite{mophoswarmRepo2022}.

\section*{Supplementary Materials}
Exoskeletons CAD files \\
Code Phototaxis  \\
Movies S1 and S2\\
Materials and Methods\\
Supplementary Text\\
Figs. S1, S2, S3\\
Algorithm 1

\bibliographystyle{apsrev4-2}
\bibliography{morphoSwarm}

\section*{Materials and Methods}

\subsection*{Exoskeleton design and manufacturing}

\subsubsection*{Exoskeleton Manufacturing}
Exoskeletons were 3D printed using either one of the following 3D printers and materials to give similar results: {\it startAsys Objet350 Connex3} printer, using {\it veroClear} (modulus of elasticity $E=1-2\;GPa$, density $\rho = 1.2\;g/cm^3$),  {\it 3D systems projet 2500plus} printer using {\it VisiJet M2R-CL (MJP)} (density $\rho = 1.14\;g/cm^3$, Elastic modulus $E\approx 1\;GPa$), {\it Prusa i3} printer using {\it PLA} (density $\rho = 1.2\;g/cm^3$, elastic modulues $E=5\;GPa$).

\subsection*{Data acquisition and experimental setups}

\subsubsection*{Phototaxis experiments}
Phototaxis experiments were performed on a $5\;\rm{mm}$ translucent plexiglass sheet placed on the floor (PMMA  27000 Diffusant Laiteux 40\%, WEBER METAUX). A circular region 6\% of the arena was lit using a projector (EPSON EB-1795F), at  RGB = (10,255,255), at approximately the spectral sensitivity peak (570 nm) of the Kilobot's light sensor (TEPT5700, Vishay Semiconductors). To reduce interference of ambient light with the light sensor, the arena was lit using four desk lamps directed at the surrounding walls (for a homogeneous diffusive illumination) and covered with red cellophane (where the robot's light sensor's sensitivity is at a minimum). Red cellophane also covered the camera used for acquisition (PIXELINK.PL-D734MU), to reduce the saturation from the lit region, and imaging of the robots. Image acquisition was carried out using $\mu$icromanager \cite{Edelstein2014}, at 2 frames per second. Robots are counted when fully in the light and results shown in Figs.~\ref{fig:morphoswarm}F,~\ref{fig:Learning}G are the average of four realizations with error bars being the standard deviation.

\subsubsection*{Code used}
Robots were programmed to phototaxis using variations of the following two codes: 1. {\bf phototaxisAdhoc.c} --- robots stand when their photoperceptron is activated given a predefined threshold. 2. {\bf phototaxisLearning.c} --- Robot stand when photoperceptron is activated but light threshold are randomly initialized and evolve through the learning algorithm.

\subsubsection*{Inclined plane experiments}
Coupling between external force and robot's orientation was measured by letting robots run on a plane made of $5\;\rm{mm}$ thick plexiglass placed at an incline, and imaged using SonyAlphaS7 camera, with video-rate image acquisition at 30 frames per second. See supporting Movies 3-6.

\subsubsection*{Data analysis}
Raw images were preprocess using ImageJ \cite{Rueden2017}, or FFmpeg, followed by particle locating using trackpy package \cite{Allan2019}, or custom code, and trajectories were linked using trackpy \cite{Allan2019}.

\subsubsection*{Simulating Morphobots}
Computer simulations were carried out by adding a stochastic term to the Eqs. \ref{eqVelocityDif}, \ref{eqOrientationDif} and using 5th order Runge-Kutta for the time propagation of Brownian dynamics of active soft-discs by adopting code from previous publications~\cite{Oppenheimer2019,Oppenheimer2022,BenZion2022a}.

\subsubsection*{Custom Morphobots}
Design files and bill of materials for custom Morphobots (Fig. \ref{fig:seesaw} in the main text) are found in the extended data. At their core, custom Morphobots are made from a 3D-printed flat chassis onto which a battery (LIR 2477), an electric switch, and two vibration motors (BestTong 14kRPM 3 Volts) are mounted. The chassis has two holes for mounting flexible curved legs (which can be 3D printed separately).

\clearpage


\title{Morphological computation and decentralized learning in a swarm of sterically interacting robots --- Supplementary Materials}
\maketitle

\setcounter{figure}{0}
\renewcommand{\figurename}{Fig.}
\renewcommand{\thefigure}{S\arabic{figure}}

\subsection*{Exoskeleton design}
Exoskeletons were designed as a round chassis (diameter $d=4.8\;\rm{cm}$) with an off-center cavity where the Kilobot is pressed in. The exoskeleton stands on three, $L = 2\;\rm{cm}$ tall legs --- one leg is stiff and round, and two opposing legs that are flexible and with a rectangular cross-section (thickness $T=1\;\rm{mm}$, and width $W=0.6\;\rm{cm}$, see Fig. \ref{fig:Morphobot} in the main text as well as previous work on elastic beams~\cite{BenZion2017,Zhu2021}). The flexible legs were designed to have their first natural vibration near resonance with the vibration motors frequency (Pololu $10X2.0\;\rm{mm}$ vibration motors,  $f_{motor} \approx 250\;\rm{Hz}$). This was done by estimating the legs to be elastic beams which are described dynamically using the Euler–Bernoulli beam theory for a homogeneous thin beam 
%
\begin{equation}
EI \frac{\partial^4 h}{\partial x^4} = -\mu\frac{\partial^2 h}{\partial t^2},
\label{eqBeam}
\end{equation}
where $h$ is the deflection, $E$ is the elastic modulus, $\mu$ is the linear mass density (given by  $\mu = TW\rho$), and $I$ is the beam moment of inertia (given by $I=\frac{W \times T^3}{12}$). In Fourier space, Eq. \ref{eqBeam} becomes
\begin{equation}
\omega = \sqrt{\frac{EI}{\mu}} k^2,
\label{eqBeamFourier}
\end{equation}
where $\omega$ is the angular frequency, and $k$ the wave number. The wavelength of the first harmonic $\lambda_1$ of a beam of length $L$ fixed at one end is $\lambda_1 = 4L$, along with Eq. \ref{eqBeamFourier} allows us to find that the natural first harmonic of the leg is to leading order 
\begin{equation}
f_1=\frac{\pi}{16\sqrt{6}}\sqrt{\frac{E}{\rho}}\frac{T}{L^2}.
\label{eqLegFrequency}
\end{equation}
Note that Eq.~\ref{eqLegFrequency} is independent of the width. Given the materials and geometry used (see the previous section), $f_1\approx 250\;\rm{Hz} \approx f_{motor}$. 

\subsection*{Individual robot dynamics}
\subsubsection*{Speed of an individual persistent particle}

The effective diffusion constant of a run-and-tumble particle, that covers a distance $l_p$, during the time interval $\tau_{RT}$ is $D_{eff} \approx \frac{1}{4} l_p^2/\tau_{RT}$ \cite{Howse2007}.  In the run and tumble mode, each Morphobot is programmed to run for $\tau_{run} = 2\;\rm{s}$ at its nominal speed of $v_0\approx5\;\rm{cm}/\rm{s}$, then perform a tumble where it pivots on average for $\tau_{pivot}=4\;\rm{s}$. The step size is thus $l_p=v\cdot \tau_{run}\approx10\;\rm{cm}$, and the total  duration is $\tau_{RT} = 6\;\rm{s}$. The resulting effective diffusion constant is $D_{eff}\approx \frac{1}{4} \left(10\;\rm{cm}\right)^2/6\;\rm{s}=4\;\rm{cm}^2/\rm{s}$.

\subsubsection*{Particle orientation as a function of trajectory length under constant force}\label{secOrientedParticle}
The dynamics of the orientation, $\theta$, relative to the positive $\hat{x}$ direction, for an active particle in the over-damped limit in the absence of noise is found to be that of a simple overdamped pendulum (Eq.  \ref{eqPendulum} in the main text)~\cite{BenZion2020}:
\begin{equation} 
\dot{\theta} = -\kappa\mu f\rm{sin}\theta.
\label{eqSimplePendulum}
\end{equation}
It is helpful to formulate Eq. \ref{eqSimplePendulum} as a function of $s$, the trajectory traveled. In 2D, $s$ is defined locally by the Euclidean distance:
\begin{equation}
ds^2 = dx^2 + dy^2.
\label{eqEuclidean}
\end{equation}
Dividing Eq. \ref{eqEuclidean} by $dt^2$ reminds us that 
\begin{equation}
\left(\frac{ds}{dt}\right)^2= \left(\frac{dx}{dt}\right)^2+\left(\frac{dy}{dt}\right)^2=v^2 
\label{eqSpeedArclength}
\end{equation}
We can find $v^2$ by multiplying Eq. \ref{eqVelocityDif} (main text) by itself:
\begin{equation}
v^2 = v_0^2 + \left(\mu f\right)^2 + v_0\mu f\rm{cos}\theta.
\label{eqSpeedMagnitude}
\end{equation}
Using the chain rule, $\frac{d\theta}{dt}=\frac{d\theta}{ds}\frac{ds}{dt}$, and plugging into Eqs. \ref{eqSpeedArclength} and  \ref{eqSpeedMagnitude}, we get an ODE describing $\theta$ as a function of $s$, the arclength:
\begin{equation}
\frac{d\theta}{ds}=\frac{\kappa f\rm{cos}\theta}{\sqrt{v_0^2 +f^2+v_0 f\rm{cos}\theta}}.
\label{eqDifArclengthTheta}
\end{equation}
Equation \ref{eqDifArclengthTheta} can be solved exactly, however, it is more instructive to examine the limit of $\frac{v_0}{\mu f}\ll1$ which then simplifies to
\begin{equation}
\frac{d\theta}{ds}=\kappa\rm{cos}\theta.
\label{eqDifArclengthThetaSimple}
\end{equation}
For $\theta\left(s=0\right)=\frac{\pi}{2}$, Eq. \ref{eqDifArclengthThetaSimple} gives Eq. \ref{eqOrientation} in the main text:
\begin{equation}
\theta\left(s\right) = 2\rm{atan}\left(e^{-\kappa s}\right).
\label{eqOrientationArclength}
\end{equation}

\subsection*{Phototaxis without learning}

\subsubsection*{Phototaxis Arena}

\begin{figure}[htbp]
\centering
\includegraphics[width=0.80\columnwidth]{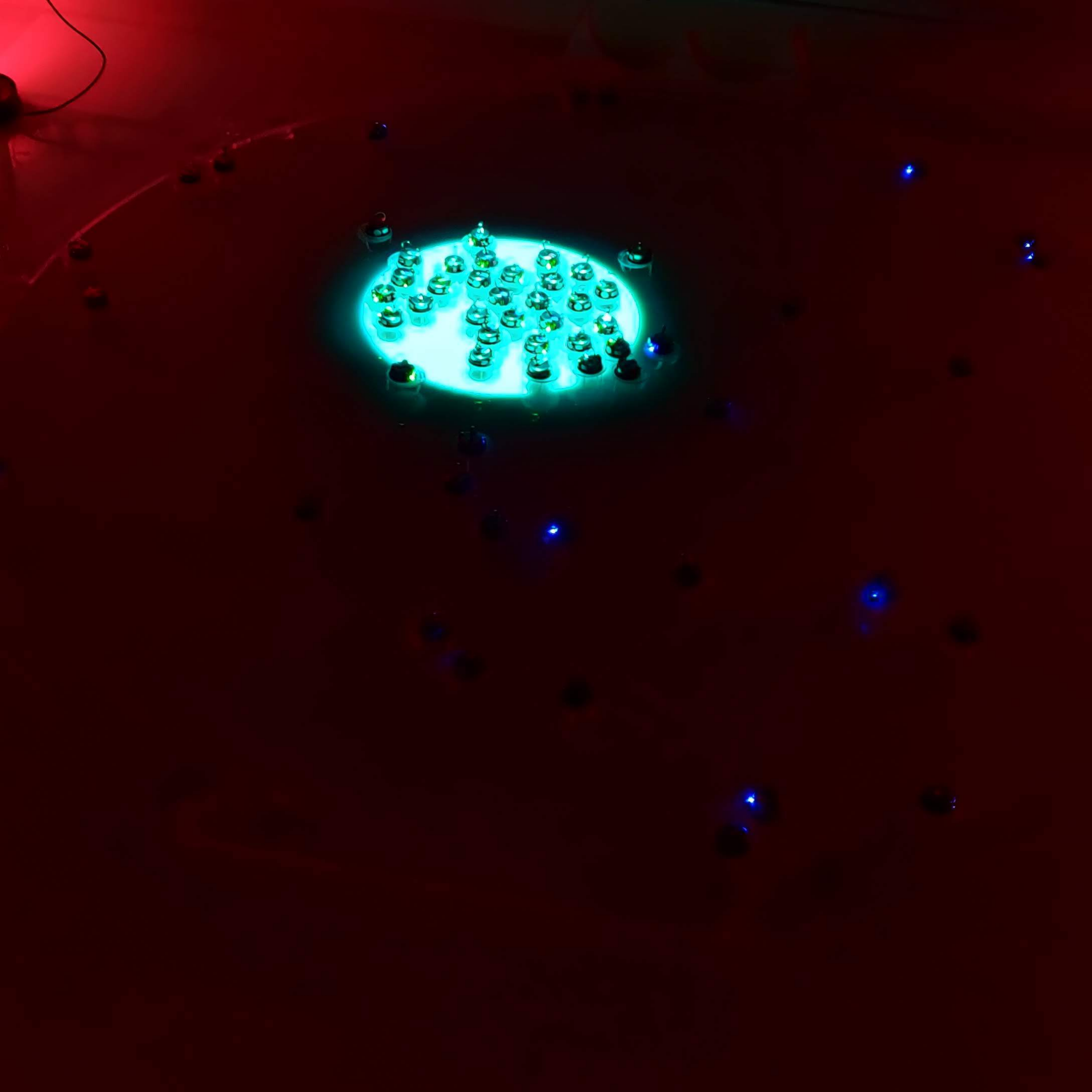} 
\label{fig:arena}
\caption{Phototaxis arena.}
\end{figure}

\subsubsection*{Photoperceptron}
Each robot has a photoperceptron encoded to measure environmental cues, and take a specific action given whether the perceptron fired or not. The minimal architecture for the photoperceptron implemented has a neural network of a single neuron (perceptron), with two inputs, bias $b = x_0$, and the locally measure light power $p = x_1$, and their associated weights $w_0$ and $w_1$, which together effectively define the threshold for firing. The perceptron output is $p$, a logistic function of the form 
\begin{equation}
p = \frac{1}{1+e^{-h}}.
\label{eqPhotoperceptron}
\end{equation}
where $h=\sum_{i=0}^{1}w_ix_i$

\subsubsection*{Sense-act cycle}\label{sec:senseAct}
The sense-act cycle uses the robot's local light intensity measurement, where the robot chooses between two meta-behaviors: {\it run} and tumble or {\it walk} and tumble, given the photoperceptron is below threshold $P<0.5$, or above threshold $P\ge0.5$, respectively. In both cases, the motion is a sequence of a moving phase (run/walk) and a tumble phase.  In the tumble phase, the robot randomly chooses a direction (clockwise or counterclockwise) and a random duration with a mean of 6 seconds. Randomizing the duration compensates for the robots' internal dynamical variability and environmental variability. For example, robots with stronger/weaker motors will take less/more time to tumble and complete the same reorientation. Similarly, a robot near a wall (or another robot) or in solitude. When in {\it run} mode, the robot sets both motor values to high, approximately moving at a straight line (persistence length much larger than the robot's size), at its nominal speed $V_0=v_0$. When walking, the robot moves for a duration $\tau_{\rm{walk}}$, and then stands for a duration $\tau_{\rm{stand}}$. The ratio between the two controls the mean speed of the robot when walking $V_1 = \frac{\tau_{\rm{walk}}}{\tau_{\rm{walk}}+\tau_{\rm{stand}}} V_0$. When $\tau_{\rm{walk}}=0$ the robot stands when the photoperceptron fires.

\subsubsection*{Phototaxis rate for low concentration noninteracting robots estimated using first arrival time}
\label{sec:Phototaxisrate}
The rate of robot arrival to the light spot at early times (where the lit region is mostly empty), can be found from the reaction rate constant, $k$, using Smoluchowski first arrival given a number concentration $c$. In 2D the mean first arrival time, $\tau_{FA}=1/kc$, of a Brownian particles with diffusion constant $D$, arriving at a target of size $b$ in an arena of diameter $d$ is given by

\begin{equation}
k = \frac{2\pi D}{\rm{log}\frac{d}{b}}.
\label{eqFirstArrival}
\end{equation}

The rate at which robots are subtracted from the dark region $\frac{dN_{bulk}}{dt}$, is given from a first-order reaction rate

\begin{equation}
\frac{dN_{bulk}}{dt} = -k N_{bulk},
\label{eqFirstOrder}
\end{equation}

which solves readily to $N_{bulk} = A\rm{exp}\{-kt\}$. Given an initial number of robots, $N_0$, the number of robots in the light, $N$, is given from the robot conservation $N_0 = N+N_{bulk}$,

\begin{equation}
N\left(t\right) = N_0\left(1-e^{-kt}\right),
\label{eqDifLimPT}
\end{equation}
which is plotted in Fig. \ref{fig:morphoswarm}.

\subsubsection*{In-silico results of phototactic swarms}
\begin{figure}[htbp]
\centering
\includegraphics[width=1\columnwidth]{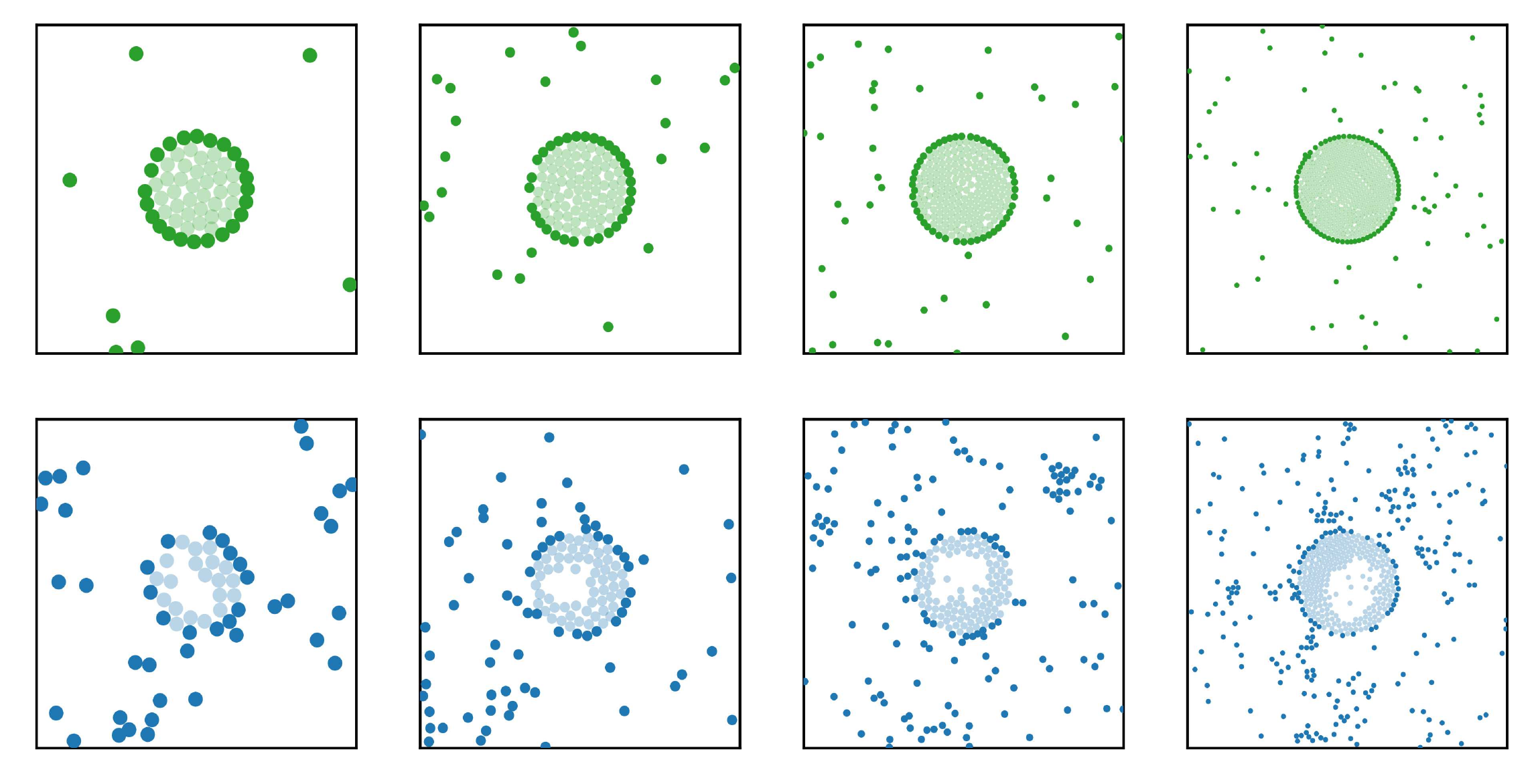} 

\caption{Snapshots from the phototaxis process of ad-hoc calibrated agents in swarms of size 64, 128, 256, 512 (left to right) of fronters (top) and aligners (bottom). Aligners tend to stop at the perimeter and leave a hollow while swarms of fronters display a more cooperative phototaxis where robots push already taxied robots deeper into the lit region}
\label{fig:simulationSnapshotsAdHoc}
\end{figure}

\begin{figure}[htbp]
\centering
\includegraphics[width=1\columnwidth]{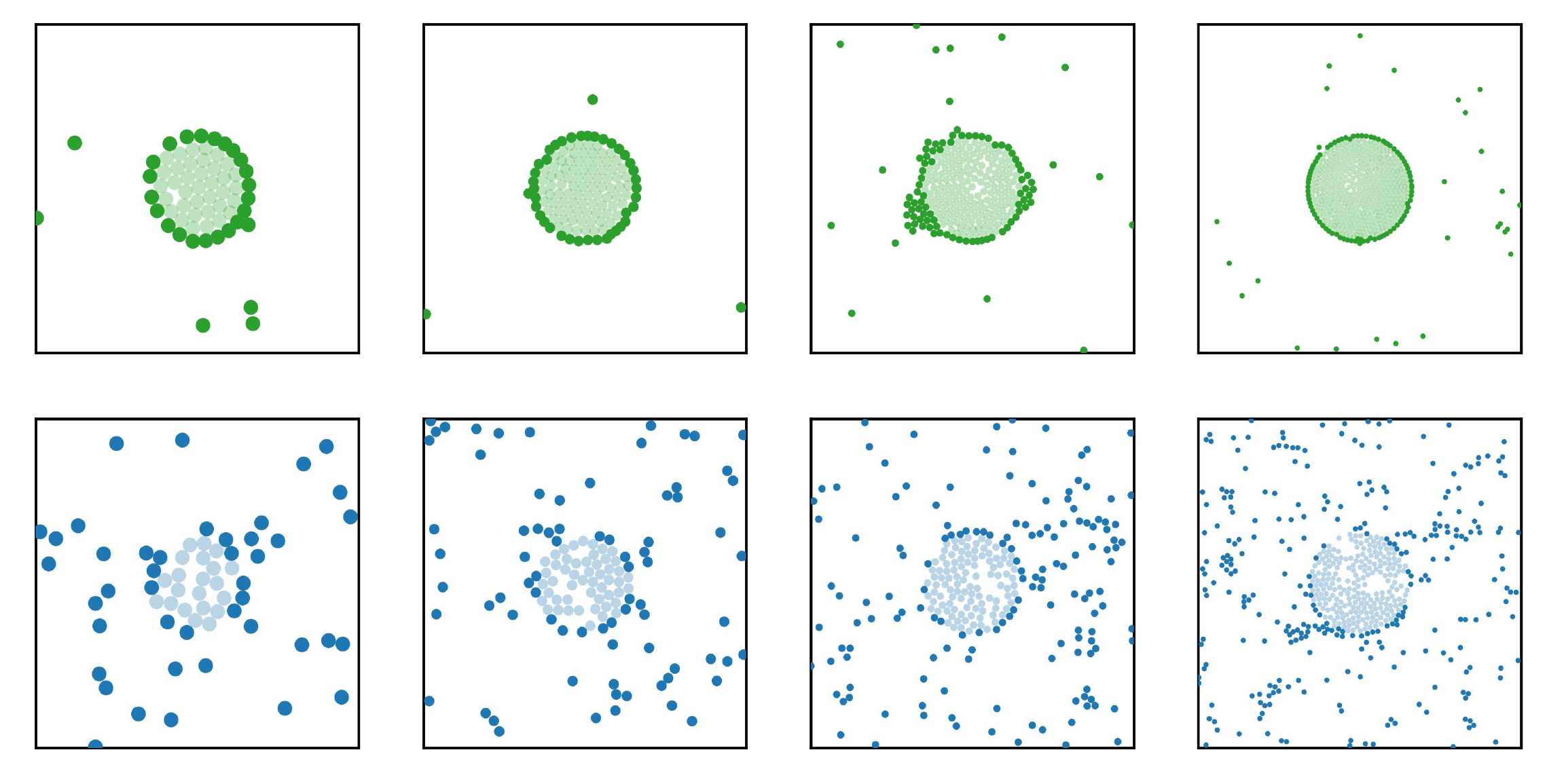} 

\caption{Snapshots from the learning process of simulated phototactic swarms of size 64, 128, 256, 512 (left to right) of fronter (top) and aligners (bottom). Swarms of aligners are arranged more loosely in the lit region relative to the tightly packed swarms of fronters.}
\label{fig:simulationSnapshotsLearning}
\end{figure}

The spatial distribution of a phototactic swarm of aligners and fronters becomes increasingly distinct with increasing swarm size. Figure~\ref{fig:simulationSnapshotsAdHoc} shows snapshots from simulations of the ad-hoc calibrated phototactic swarm. Showing that fronters fill the lit region, while aligners develop a ring at the perimeter of the light spot, and that the effect grows with swarm size. The effect of morphology becomes more pronounced with growing swarm size also for learning swarms as can be seen in Fig.~\ref{fig:simulationSnapshotsLearning}.

\subsection*{Collective learning in a sterically interacting swarm}

In this final section, we derive an analytic link between the behavior of individuals in a swarm and the convergence of a decentralized learning scheme. We first establish empirically that a swarm where robots are allowed to collide has good mixing, a property required to establish individual-to-collective correspondence. We then cast collective learning as a dynamical process where robots in the swarm can exchange their state upon encounter. And finally, we show that given the reward function, $\rho$, as defined in the main text (Eq. \ref{eqReward}), a randomly initialized swarm will converge on a successful phototaxis strategy.

\subsubsection*{Individual dynamics of pre-calibrated robots}
\begin{figure*}[t]
\centering
\includegraphics[width=0.85\textwidth]{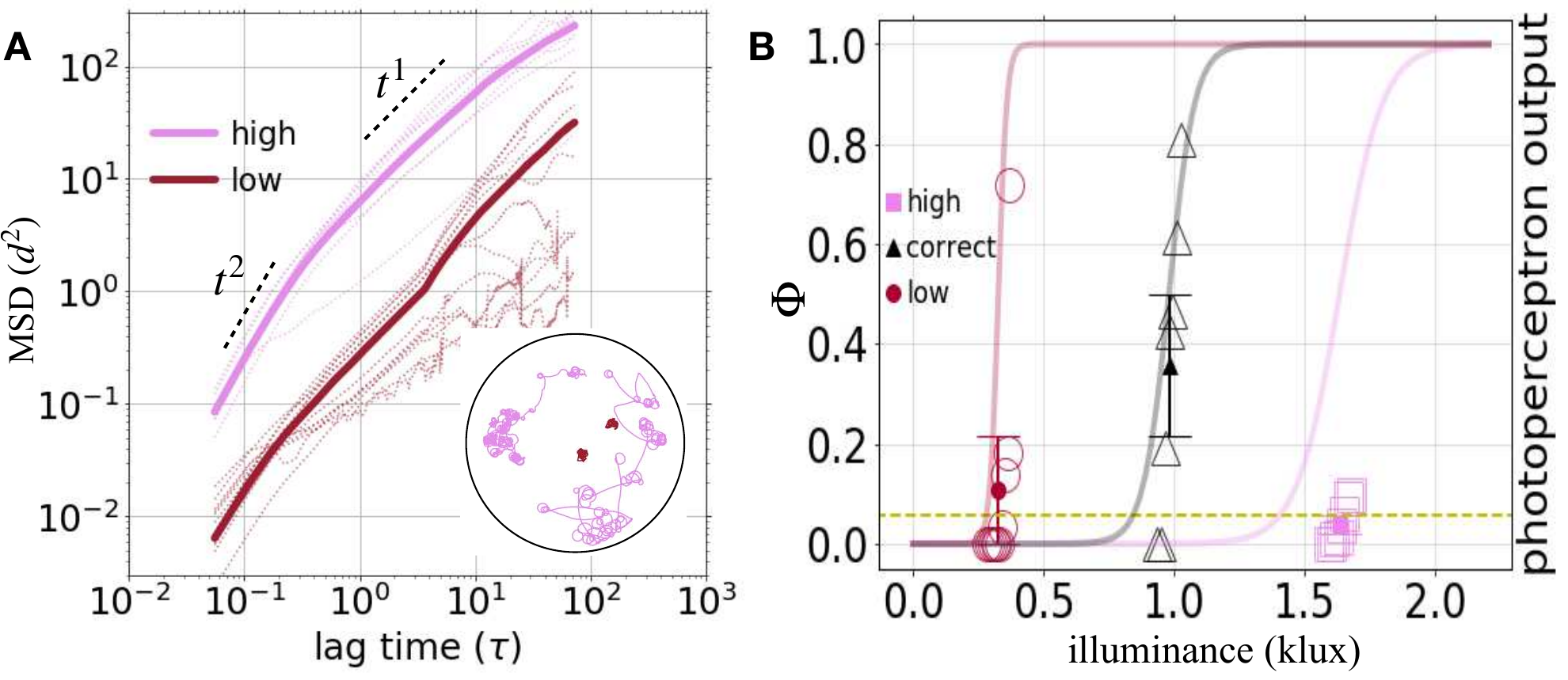} 
\caption{Dynamics and performance of individual pre-programmed robots. {\bf A} Mean square displacement of the two meta behaviors: the fast run-and-tumble (pink), or the much slower walk-and-tumble (red), with a fixed speed ratio between run and walk. The run-and-tumble behavior transitions from ballistic ($\propto t^2$) to linear ($\propto t^1$) at approximately one persistent time $\tau\equiv t/\tau_p\approx 1$. Inset shows typical, 20 minutes long trajectories of both run-and-tumble (pink) and walk-and-tumble (red) inside the circular 150 cm wide arena. {\bf B} The time fraction in the light $\Phi$ of an individual robot shows that a robot with too high or too low light threshold performs as well as random, while a robot that has a correct policy, is responsive to the light source. The policy of each robot is determined by the weights of the photo-perceptron, setting the light threshold to activate the photo-perceptron (solid curves).}
\label{fig:IndividualDynamics}
\end{figure*}

The mean time a robot spends in the light spot, $\Phi$, depends on three quantities: 1. The ratio between the speed of the robot in the light $V_1$ to its speed outside the light, $V_0$; 2. The area fraction of the light spot, $\sigma$; 3. Whether or not the robot has the correct threshold for phototaxis. Allowing individual, pre-set robots in the light patterned arena to perform phototaxis shows two distinct behaviors: robots with a correct light threshold for phototaxis spend more time in the lit region than robots with a wrong threshold (either too high or too low, see Fig. \ref{fig:IndividualDynamics}B) for which time in the light is as good as random $\Phi_W \approx \sigma$.

\subsubsection*{Collective dynamics of a pre-calibrated swarm}

\begin{figure*}[t]
\centering
\includegraphics[width=0.85\textwidth]{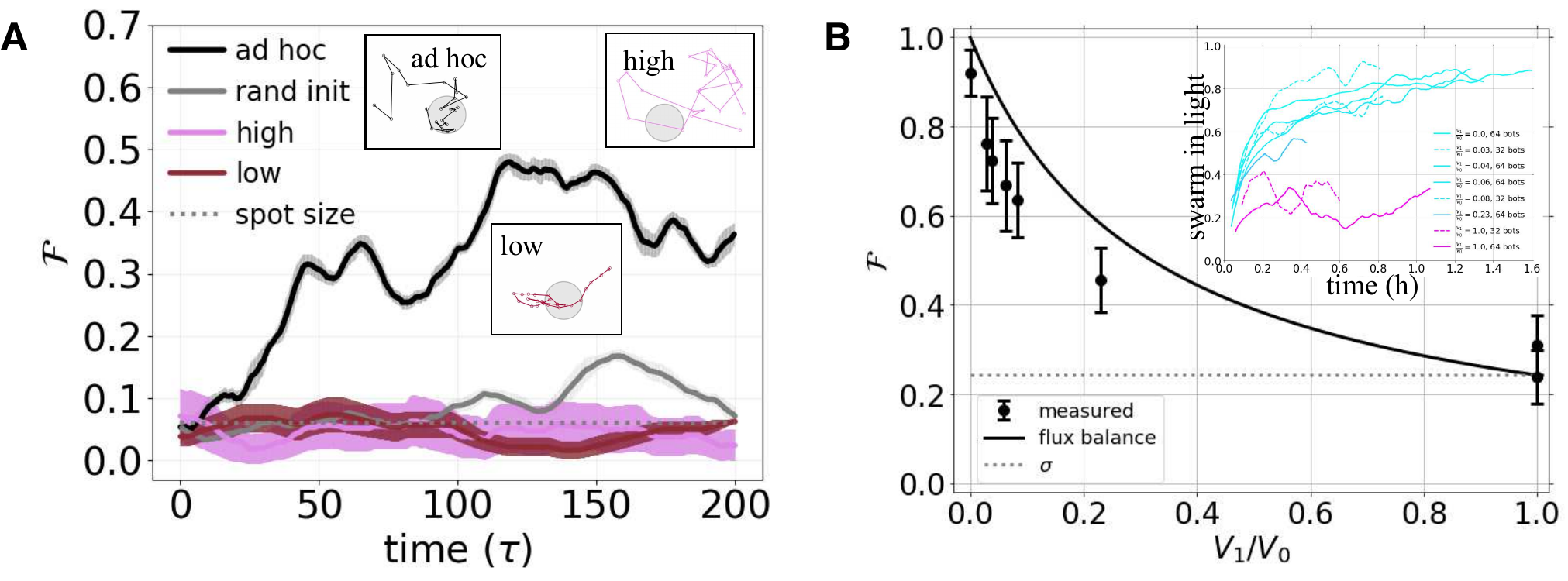} 

\caption{Collective performances of pre-programmed robots.  {\bf A} Swarm fraction in the light for a fixed speeds ratio ($V_1/V_0 = 1/15$)  different initializations: correctly calibrated swarm (black), low (red) high (pink) threshold, and randomly initialized swarm (gray). Insets show typical trajectories showing both low and high are agnostic to the light spot, while the correctly, ad-hoc, initialized robot, dwells longer in the light spot, increasing its reward $\rho$. {\bf B} The run-to-walk speed ratios determine the collective ability of a swarm to phototaxis, and a steady-state flux balance model (see Eq. \ref{eqSwarmFraction}) quantitatively predicts the percentage of the swarm in the light, with no fitting parameters.}
\label{fig:LargeLightSpot}
\end{figure*}

To test how the speed ratio $V_1/V_0$ affects the number of bots in the light, phototaxis experiments were done by increasing the size of the lit region to 25\% of the arena. Experiments were performed with either $64$ bots or $32$ bots (see inset in Fig. \ref{fig:LargeLightSpot}B). The speed ratio was set by controlling the $\tau_{\rm{stand}}$ as described in Section \ref{sec:senseAct}, changing the speed from standing in light $V_1/V_0 = 0$ to effectively being agnostic to the light $V_1/V_0=1$ (see Fig.~\ref{fig:LargeLightSpot}B). The relative number of robots in the light was tracked over up to 2 hours, and the mean and standard deviation of the last $10\%$ of the experiment duration were used to evaluate the steady-state fraction of the swarm in the light, $\mathcal{F}_{ss}$.

\subsubsection*{Collective decentralized learning}

The collective learning was implemented by deploying a randomly initialized swarm of Morphobots that follow the algorithm detailed below (see Algorithm \ref{algoVanillaHIT}) to give the results presented in Fig.\ref{fig:Learning} in the main text. 

\begin{algorithm}[ht!]
	\SetAlgoLined
	\KwData{\\
	    $i$ : the unique identifier of the current robot,\\
		$T$ : evaluation time,\\
		$\pi$ : Policy function,\\
		$\boldsymbol{w^i}$ : Random uniform initialisation of policy parameters\\
		$R[T]$ : Empty reward buffer of size $T$,\\
		$r$ : Current light intensity,\\
		$\rho^i$ : performance self-assessment during $T$,\\
		$\boldsymbol{\mathrm{a}}$ : Null action vector,\\
		$\boldsymbol{\mathrm{o}}$ : Null observation vector\\
	}
	\Begin{
		$t = 0$\\
	   	 \While{True}{
	    		$\boldsymbol{\mathrm{o}}, r$ = sense()\\
		        $R[t\mod T]=r$\\
		        $\boldsymbol{\mathrm{a}} = \pi(\boldsymbol{\mathrm{o}}|	\boldsymbol{w^i}$)\\
	        act($\boldsymbol{\mathrm{a}}$)\\
		        \If{$t > T$}{
   			         $\rho^i = \sum_{k=0}^{T-1}R[k]$\\
				broadcast($\boldsymbol{w^i}$,$\rho^i$)\\
		        	 	\If{$new\_message$}
				{
            				$\boldsymbol{w^j}$,$\rho^j$ = $decode\_message$\\
                		       			\If{$\rho^j > \rho^i$}{
                	        					$w^i = w^j$
						}
				}
			}
		$t = t + 1$
		}
	}
\caption{The phototactic-HIT algorithm}
\label{algoVanillaHIT}
\end{algorithm}

Each robot in the population runs Algorithm~\ref{algoVanillaHIT}. This Algorithm builds from the HIT algorithm originally presented in~\cite{fontbonne2020,Yoones2022}, with two important modifications. Firstly, the algorithm is designed with phototaxis as an objective (the $\rho$ objective function on line 9 measures the amount of light measured in the last $T$ time steps). Secondly, transfer of \textbf{all} control parameters is performed, which implies that whenever a robot receives a parameter set that is deemed to be performing better than its own, this parameter set is fully copied and overwrites existing parameter values. Mutation is not used in this implementation, as the number of robots used ($64$) and the low dimensionality of the search space ($\mathbb{R}^2$) allows for a sufficient amount of behavioral diversity in the initial population of robots (see~\cite{fontbonne2020} for a comprehensive study of transfer and mutation rate and operators). 

It should also be noted that the performance assessment ($\rho^i$) is not reset to zero after a parameter update, which implies that the performance assessment of a newly updated policy is under-estimated for $T$ time steps. While this may temporarily lead to theoretically best-performing control parameters being overwritten before $T$ steps elapsed, the elitist selection scheme ensures that the original best-performing individual is never lost.

Finally, note that in the algorithm the observable $o$ contains information about the current light intensity $r$. They are separated for semantic reasons as $o$ is used as an input value for the policy $\pi$, while $r$ is used to update the performance self-assessment $\rho^i$.

\subsubsection*{Collective learning as a dynamical process}
The goal is to link policy and environment to find a route for a successful decentralized learning scheme on average. 
The time evolution of the probability $p^i_C$, respectively $p^i_W$, of finding a robot $i$ with a correct, respectively wrong, policy is described by the following master equation~:
\begin{equation}
\frac{dp^i_C}{dt} = \omega^i_{C,W} p^i_W- \omega^i_{W,C} p^i_C,
\label{eqMasterReduced}
\end{equation}
where $p^i_W = 1-p^i_C$ and $\omega^i_{C,W}$, respectively $\omega^i_{W,C}$, are the transition rates at which the robot $i$ with a wrong policy turns into a correct policy and conversely. 
Defining the concentration of robots with a correct policy as $c_C$, the concentration of robots with a wrong policy is given by $c_W = c_0 - c_C$ (where $c_0$ is the constant overall concentration of the swarm). In a system with good mixing, the encounter rate between robots with correct and wrong policy is proportional to the product of their concentrations, $k c_Cc_W$ (to leading order), where $k$ is proportional to mutual diffusion. When encountered, robots compare their rewards and a robot with a wrong policy will adopt the correct policy at a probability $f_{CW}$. The opposite is also possible (false information scenario) and a robot with a correct policy will adopt a wrong policy at a rate $f_{WC}$. When combined, the rate equation for the growth in the population with a correct policy becomes 
\begin{equation}
\frac{dc_C}{dt} = k \left(f_{CW}-f_{WC}\right) c_C c_W.
\label{eqLearningRateMain}
\end{equation}

Collective learning is achieved when the concentration of robots with correct policy grows, implying: $dc_C/dt >0$. Since $k$, $c_W$, and $c_C$ are all non-negative, collective learning is guaranteed when $f_{CW}>f_{WC}$. Given the definition of the reward function in Eq.~\ref{eqReward} and the selection protocol defined in Fig.~\ref{fig:Learning}, robots with the correct threshold will have, on average, a higher reward, $\rho_C>\rho_W$, and the population with the correct policies will prevail (see SM for a full derivation).

We assume space is homogeneous and identify the probability of a given policy for a given robot with the concentration of robots, $c_{C,W}$, having that policy. Since transition between states happens only through local communication, the transition rates are proportional to the robot encounter rate constant, the latter being independent of the policies of the encountering robots. Under the above hypothesis, the transition rates are proportional to the concentrations and the transition elements read $\omega_{C,W}\approx k f_{CW} c_C$, and symmetrically $\omega_{W,C}\approx k f_{WC} c_W$, where $k$ is the encounter rate, and $f_{CW}$, respectively $f_{WC}$, is the transfer probability of $W$, respectively $C$, becoming $C$, respectively $W$, upon encounter. As a result,
\begin{equation}
\frac{dc_C}{dt} = k \left(f_{CW}-f_{WC}\right) c_C (c_0-c_C),
\label{eqLearningRate}
\end{equation}

where $c_0$ is the overall concentration of robots in the arena.

To achieve positive mean collective learning, $dc_C/dt>0$, the factors of the product in Eq. \ref{eqLearningRate} has to be non negative, which amounts to

$$ f_{WC}>f_{CW}$$

As learning is decentralized, $f$ is an embedded function and can depend only on information locally accessible to the robot. This includes the robot's policy (here, the weights of its neural network, $\{w_i\}$), its measurement history, and the message received. Here we exclude a direct functional dependence of $f$ upon the policy as it will artificially bias the learning to a given trait (say turn left) instead of the desired fitness (say measure light). \footnote{Note that using the policy directly could also be useful if two robots have similar rewards but one has a superior policy (say more energetically efficient). This can also be directly introduced into the reward.} Therefore $f$ will depend on some function of the history of its own sensor input and the history of an encountered robot. In our minimal system, the measurement history contains the $M$ most recent light intensity measurements, $I_{i\in1..M}$  (stored as a rolling buffer, i.e FIFO).\footnote{In our case $I$ is an array of bytes and $M=250$.} Since the robots have a finite bandwidth, to allow asynchronous performance, where communication does not slow down dynamics, the message should be kept succinct. This means robots should only exchange a reduced version of their memory that can be contained within a message.\footnote{On the Kilobot platform, the message size is 8 bytes.} This will be a function of the memory and will be called the reward function,  $\rho=\rho\left(\{I_i\}\right)$.\footnote{In our case, $\rho$ will be a single byte.} In our implementation scheme, robots continuously broadcast a message with their reward and policy; a robot receiving a message compares the $B$roadcasted reward, $\rho_B$, to its self-reward, $\rho_S$, and if $\rho_B>\rho_S$ the policy will be {\it deterministically} adopted. The question then becomes what functional form does the local $\rho$ take, to satisfy global positive learning {\it on average}: $\langle \rho_C\rangle >\langle\rho_W\rangle$.

Empirically we found that in an ad-hoc calibrated swarm at steady state, individual robots with a correct policy spend more time in the light $\Phi_C>\Phi_W$ (see Fig.\ref{fig:IndividualDynamics}), where the mean time in the light of an individual robot is measured by $\Phi=\frac{1}{T}\int_{-T}^0dt I\left(t\right)$. Identifying the reward function as proportional to the mean time in the light, $\rho\propto\Phi$ allows us to bias the swarm towards learning the correct policy. We do this by computing the reward as the average of the light intensity buffer, $\rho = \frac{1}{M}\sum_{i=1}^MI_i$ (Eq.~\ref{eqReward} in the main text).

Recalling that the transfer probability function depends only on the received and embedded reward, the condition for learning simply becomes that on average, correct robots have a greater reward:
\begin{equation}
\rho_C >\rho_W
\label{eqLearningCondition}
\end{equation}

\subsubsection*{Mean reward in a well-mixed swarm}
To satisfy the learning condition presented in the previous section (Eq. \ref{eqLearningCondition}) we recall the good mixing property found empirically in previous sections wherein the locally computed reward can approximate an individual's time average, which is a proxy to the global performance $\rho \propto \Phi \approx \mathcal{F}$~\cite{BenZion2021c}. A swarm with good mixing can be modeled as a fluid that obeys the continuity equation:

\begin{equation}
\frac{d C}{dt} = -\nabla J
\label{eqContinuity}
\end{equation}
where $C$ is concentration and $J$ is the flux. At steady-state, the concentration does not change in time, and so the left-hand side of Eq. \ref{eqContinuity} vanishes $\frac{dC}{dt} = 0$. In a system made of a collective with spatially varying speed, it has been shown that concentrations are inversely proportional to speeds (see Schnitzer who studied field taxi in bacteria~\cite{Schnitzer1993}). For the simple case discussed here, space is divided into two regions, lit and dark, where robots have speeds $V_1$ and $V_0$ respectively. This results in the following relation of concentration of robots in the light given their speed in the light,

\begin{equation}
C_0 V_0 = C_1 V_1,
\label{eqCVCV}
\end{equation}
where $C_1$ and $C_0$ are the concentrations of the robots in the light and in the dark respectively. We next derive a link between Eq. \ref{eqCVCV} and the definition of the swarm fraction in the light, $\mathcal{F}$ as defined in Eq. \ref{eqSwarmFraction} the main text. The concentration of robots in the light, $C_1$ is the number of robots in the light over the area of the lit region $C_1 = N_1/\sigma A$. The concentration of robots in the dark is the remaining number of robots in the swarm, $N_0=N-N_1$ over the remaining area of the arena: $N_0/(1-\sigma)A$. Plugging these into Eq. \ref{eqCVCV} and rearranging gives,

\begin{equation}
\mathcal F_{ss} = \frac{1}{\frac{V_1}{V_0}\left(\frac{1}{\sigma}-1\right)+1},
\label{eqSwarmFraction}
\end{equation}
connecting the collective performance (fraction of the swarm in the light), to individual policies (speed ratio), given the environmental condition (fraction of the arena that is lit). For the case of an agnostic swarm, $V_0=V_1$, the swarm does not perform better than random $\mathcal{F}\approx \sigma$ (consistent with measured performance of individuals Fig. \ref{fig:IndividualDynamics} and collectives Fig. \ref{fig:LargeLightSpot}). When the robots are responsive to the light, $V_1<V_0$, they accumulate in the light spot, and both the individual dynamics (Fig. \ref{fig:IndividualDynamics}B) and the collective dynamics (Fig. \ref{fig:LargeLightSpot}B), are also consistent with the above derivation with Eq. \ref{eqSwarmFraction} setting the high bound for the swarm's performance. When robots are responsive, the swarm accumulates in the light, and individual robots spend more time in the light. Since the reward function as defined in Eq. \ref{eqReward} is proportional to the time a robot spends in the light $\rho\propto \Phi$, robots that slow down in the light will satisfy the learning condition (Eq. \ref{eqLearningCondition}), and their superior policy will spread throughout the swarm.

\end{document}